%% file: main.tex
\documentclass[final,onefignum,onetabnum]{siamart190516}
\usepackage{amssymb}
\usepackage{bbm}
\usepackage{bm}
\usepackage{siunitx}

\input{shared}

\graphicspath{{figs/}}

\ifpdf
\hypersetup{
  pdftitle={Physics-informed neural networks with hard constraints for inverse design},
  pdfauthor={L. Lu, R. Pestourie, W. Yao, Z. Wang, F. Verdugo, and S. G. Johnson}
}
\fi

\begin{document}

\maketitle

\begin{abstract}
Inverse design arises in a variety of areas in engineering such as acoustic, mechanics, thermal/electronic transport, electromagnetism, and optics. Topology optimization is a major form of inverse design, where we optimize a designed geometry to achieve targeted properties and the geometry is parameterized by a density function. This optimization is challenging, because it has a very high dimensionality and is usually constrained by partial differential equations (PDEs) and additional inequalities. Here, we propose a new deep learning method---physics-informed neural networks with hard constraints (hPINNs)---for solving topology optimization. hPINN leverages the recent development of PINNs for solving PDEs, and thus does not rely on any numerical PDE solver. However, all the constraints in PINNs are soft constraints, and hence we impose hard constraints by using the penalty method and the augmented Lagrangian method. We demonstrate the effectiveness of hPINN for a holography problem in optics and a fluid problem of Stokes flow. We achieve the same objective as conventional PDE-constrained optimization methods based on adjoint methods and numerical PDE solvers, but find that the design obtained from hPINN is often simpler and smoother for problems whose solution is not unique. Moreover, the implementation of inverse design with hPINN can be easier than that of conventional methods.
\end{abstract}



\section{Introduction}
\label{sec:intro}

Metamaterials are artificial materials that achieve targeted properties through designed fine-scale geometry rather than via material properties. They arise in a variety of areas in engineering, e.g., acoustics, mechanics, thermal/electronic transport, and electromagnetism/optics~\cite{kadic20193d}. Designing a metamaterial's geometry for a particular functionality is a challenging task, because the design space has a very high dimensionality (millions to billions of parameters). In contrast to intuition-based approaches using heuristics and handful of hand-tweaked geometric parameters, \textit{inverse design} starts with the targeted functionality and sets it as an objective function to be optimized via partial-differential-equation-constrained (PDE-constrained) large-scale optimization~\cite{molesky2018inverse}. When the parameterization of the geometry is via the parameters of discrete geometric components, inverse design is called \textit{shape optimization}~\cite{pestourie2018inverse, bayati2020inverse}. When the parameterization of the geometry is via a density function or level set so that the connectivity/topology is arbitrary, inverse design is called \textit{topology optimization}~\cite{molesky2018inverse}; this is the variant of inverse design considered in this study.

PDE-constrained inverse design, especially PDE-constrained topology optimization, is important for various subjects in computational science and engineering, e.g., optics and photonics~\cite{molesky2018inverse,pestourie2018inverse,bayati2020inverse}, fluid dynamics~\cite{borrvall2003topology,guest2006topology,duan2016topology,schulz2016computational}, and solid mechanics~\cite{maute2013topology,bendsoe2013topology}. Different numerical methods have been developed to solve PDE-constrained inverse design, and these traditional methods are commonly based on a numerical PDE solver (either approximate~\cite{pestourie2018inverse} or brute force~\cite{molesky2018inverse}). The large-scale optimization is performed via gradient-based optimization algorithms, where the gradient is obtained using an adjoint method, which computes the gradient of the objective function with respect to all parameters at the cost of at most a single additional simulation.

Recently, deep learning in the form of deep neural networks has been used in many areas of inverse design, including nanophotonics~\cite{so2020deep,jiang2020deep}, mechanical materials~\cite{guo2020artificial}, aerodynamics~\cite{sekar2019inverse}, and electromagnetism~\cite{sasaki2019topology}. As the dominant approach nowadays for data-driven problems, deep neural networks are usually employed in the supervised-learning paradigm to learn the nonlinear mapping from arbitrary designs to their associated functional properties~\cite{kojima2017acceleration,peurifoy2018nanophotonic,liu2018training,white2019multiscale,sasaki2019topology,tahersima2019deep,pestourie2020assume,pestourie2020active,guo2020artificial} or vice versa~\cite{liu2018training,sekar2019inverse,tahersima2019deep}, i.e., neural networks are used as surrogate models of PDE solvers to accelerate the optimization. However, it may require a vast quantity of data to train a neural network for complex problems, and the generation of such datasets via ``brute-force'' PDE solvers could be very expensive (although many strategies are being investigated reduce the amount of training data, e.g., active learning~\cite{pestourie2020active}). Alternatively, in unsupervised learning and reinforcement learning, neural networks are employed as generators to generate candidate designs~\cite{liu2018generative,jiang2019simulator,jiang2020deep,so2020deep}, which are then evaluated by numerical solvers. Hence, all these approaches still rely on traditional numerical PDE solvers.

Physics-informed neural networks (PINNs) have been developed as an alternative method to traditional numerical PDE solvers~\cite{raissi2019physics,lu2019deepxde}. PINNs solve PDEs by minimizing a loss function constructed from the PDEs, and the PDEs are solved when the loss is close to zero. Compared to traditional numerical solvers, a PINN is mesh-free and thus can easily handle irregular-domain problems. Moreover, PINNs have been successfully employed to solve diverse \textit{inverse problems}, including in optics~\cite{chen2020physics}, fluid mechanics~\cite{raissi2020hidden,tartakovsky2020physics}, systems biology~\cite{yazdani2020systems}, biomedicine~\cite{sahli2020physics,kissas2020machine}, and even inverse problems of stochastic PDEs~\cite{zhang2019quantifying} and fractional PDEs~\cite{pang2019fpinns}.  Here, the inverse problems were to infer unknown data (such as PDE coefficients/geometries) from partial observations of the PDE solutions in some regions, and a major concern is often regularizing the problem in order to ensure a unique ``correct'' solution. In contrast, for inverse design, one may not have any data of the PDE solution, and any realizable design that approximately maximizes the objective functional is acceptable (even if it is not unique).

For conventional numerical methods, inverse problems and inverse design might be solved by similar methods, because the PDE is directly solved via a numerical solver, and we only need to find the best PDE solution to minimize the objective function or match the observed data. However, in PINN, inverse design poses new challenges. The inverse problems are solved with PINNs by using the mismatch between the PDE solution and the data as a loss function (data-based loss), so that the network is trained to minimize the sum of the data-based loss and the PDE-based loss~\cite{chen2020physics,lu2019deepxde}. Such an optimization problem can be solved relatively easily, because these two losses are consistent and can be minimized to zero simultaneously. In contrast, for inverse design, the PDE-based loss and the objective function are usually not consistent---it is not generally possible to both exactly solve the PDE and make the design objective arbitrarily good with the same solution---and so they compete with each other during optimization. Hence, a PINN that merely optimizes the sum of the objective and the PDE loss will usually end up in an optimum that is not a solution to the PDE. Moreover, inverse design problems often have additional inequality constraints, such from manufacturing constraints, that must be satisfied for the design to be acceptable.

To overcome these difficulties, we develop a new PINN method with hard constraints (hPINN) to solve PDE-constrained inverse design. We also consider inequality constraints in hPINN. We propose two approaches to impose the equality and inequality constraints as hard constraints, including the penalty method and the augmented Lagrangian method. We also use the approach of soft constraints for comparison. Imposing hard constraints to neural networks has been considered very recently~\cite{nandwani2019primal,dener2020training} for data-driven problems in the supervised learning paradigm.

The paper is organized as follows. In Section~\ref{sec:method}, after introducing the setup of inverse design and the algorithm of PINN, we present the method to exactly impose Dirichlet and periodic boundary conditions by directly modifying the neural network architecture. We then propose a soft-constraint approach and two hard-constraint approaches to impose PDEs and inequality constraints in hPINN, including the penalty method and the augmented Lagrangian method. In Section~\ref{sec:res}, we demonstrate the effectiveness and convergence of hPINN for a holography problem in optics and a fluid problem of Stokes flow. By comparing with traditional PDE-constrained optimization methods based on adjoint methods with the finite-difference frequency-domain (FDFD) method and finite element method (FEM) as the numerical PDE solvers, we find that hPINN achieves the same objective-function performance but often seems to obtain a simpler and smoother design (without imposing additional constraints on the design lengthscales or smoothness, as is typically done in topology optimization). Finally, we conclude the paper in Section \ref{sec:conc}.

\section{Methods} \label{sec:method}

We first introduce the problem setup of inverse design considered in this paper and then present the method of physics-informed neural networks (PINNs) with hard constraints (hPINNs) for solving inverse design.

\subsection{Inverse design}

We consider a physical system governed by partial differential equations (PDEs) defined on a domain $\Omega \subset \mathbb{R}^d$:
\begin{equation} \label{eq:pde}
    \mathcal{F} \left[ \mathbf{u}(\mathbf{x});\gamma(\mathbf{x}) \right] = \mathbf{0}, \quad \mathbf{x} = (x_1, x_2, \cdots, x_d) \in \Omega
\end{equation}
with suitable boundary conditions (BCs):
\begin{equation} \label{eq:bc}
    \mathcal{B}\left[ \mathbf{u}(\mathbf{x}) \right] = 0, \quad \mathbf{x} \in \partial \Omega,
\end{equation}
where $\mathcal{F}$ includes $N$ PDE operators $\{\mathcal{F}_1, \mathcal{F}_2, \cdots, \mathcal{F}_N \}$, $\mathcal{B}$ is a general from of a boundary-condition operator, and $\partial \Omega$ is the boundary of the domain $\Omega$. $\mathbf{u}(\mathbf{x}) = (u_1(\mathbf{x}), u_2(\mathbf{x}),\cdots,u_n(\mathbf{x})) \in \mathbb{R}^n$ is the solution of the PDEs and is determined by the parameter $\gamma(\mathbf{x})$, which is our quantity of interest (QoI) for the inverse design problem. ($\gamma$ generally describes some structure to be manufactured in order to realize an optimized device.)

In an inverse-design problem, we search for the best $\gamma$ by minimizing an objective function $\mathcal{J}$ that depends on $\mathbf{u}$ and $\gamma$. The pair $(\mathbf{u}, \gamma)$ must to satisfy the equality constraints enforced by the PDEs in Eq.~\eqref{eq:pde} and the BCs in Eq.~\eqref{eq:bc}; in certain situations, we may also have additional equality or inequality constraints for $\mathbf{u}$ and $\gamma$ (e.g., stemming from manufacturing constraints or multi-objective problems). In this paper, we only consider inequality constraints, as equality constraints can be tackled in the same way as the PDE and BC constraints. Then the inverse design problem is formulated as a constrained optimization problem:
\begin{equation*}
    \min_{\mathbf{u}, \gamma} \mathcal{J}(\mathbf{u}; \gamma)
\end{equation*}
subject to
\begin{equation} \label{eq:constraint}
    \left\{ \begin{array}{l} 
    \mathcal{F} \left[ \mathbf{u};\gamma \right] = \mathbf{0}, \\
    \mathcal{B}\left[ \mathbf{u} \right] = 0, \\
    h(\mathbf{u}, \gamma) \le 0,
    \end{array} \right.
\end{equation}
where the last equation is the inequality constraint(s). We note that the optimal solution $\gamma$ of this optimization problem may not be unique, and moreover there may be many acceptable local optima with similar performance. (An inverse problem could be viewed a special case of inverse design, where the objective function $\mathcal{J}$ is the error between the PDE solution $\mathbf{u}$ and the observed measurements, but in this context there is typically a unique ``ground-truth'' solution that is desired, and hence special attention must be paid to conditioning and regularization.)

\subsection{Physics-informed neural networks}

One difficulty of the constrained optimization problem is that $\mathbf{u}$ and $\gamma$ must satisfy the PDEs, and a common strategy is for $\mathbf{u}$ to be obtained from solving the PDEs for a $\gamma$ by using numerical methods such as finite differences or finite elements. In this paper, we will instead use PINNs.

\begin{figure}[htbp]
    \centering
    \includegraphics[width=\textwidth]{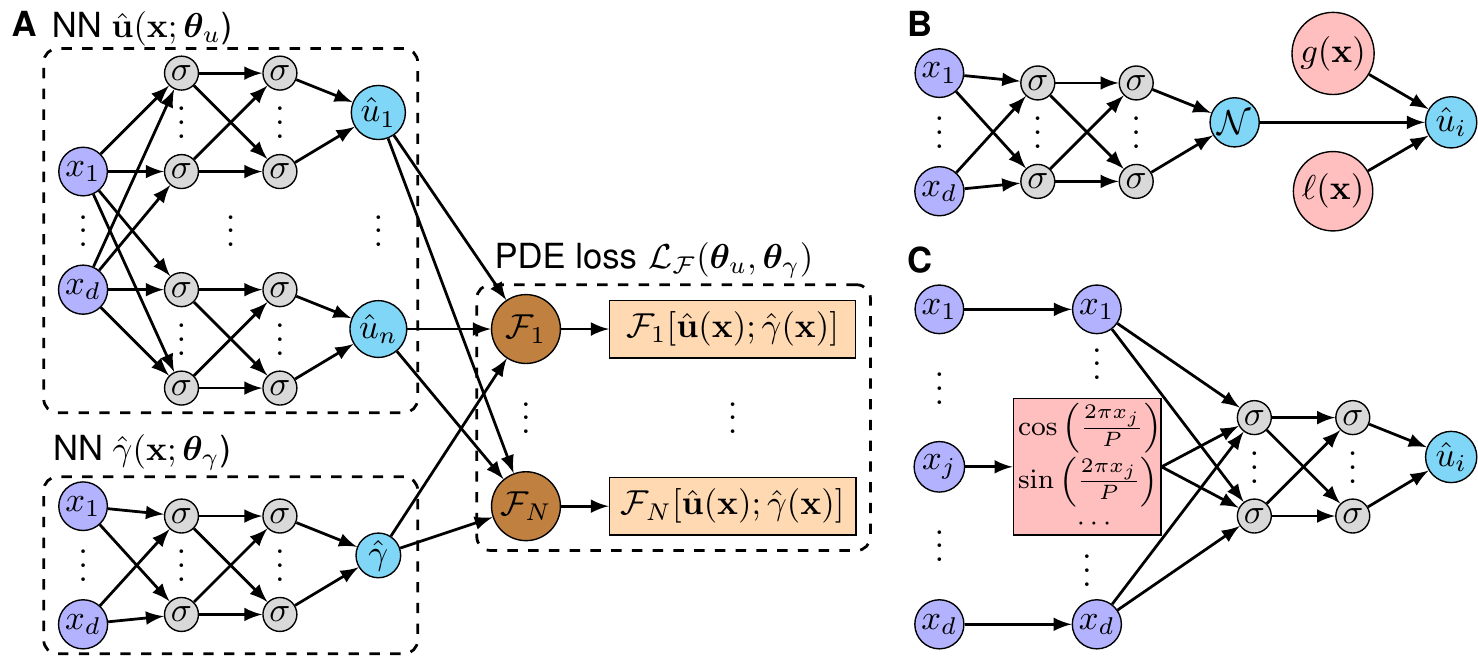}
    \caption{\textbf{Physics-informed neural networks with hard-constraint Dirichlet and periodic boundary conditions.} (\textbf{A}) Two independent neural networks $\hat{\mathbf{u}}(\mathbf{x};\bm{\theta}_u)$ and $\hat{\gamma}(\mathbf{x};\bm{\theta}_\gamma)$ are conostructed to approximate $\mathbf{u}(\mathbf{x})$ and $\gamma(\mathbf{x})$. The gradients in the PDE-informed loss is computed via AD. (\textbf{B}) Dirichlet BCs are strictly imposed into the network architecture by modifying the network output. (\textbf{C}) Periodic BCs are strictly imposed into the network architecture by modifying the network input.}
    \label{fig:pinn}
\end{figure}

In a PINN, we employ $n$ fully connected deep neural networks $\hat{\mathbf{u}}(\mathbf{x}; \bm{\theta}_u)$ to approximate the solution $\mathbf{u(\mathbf{x})}$ (Fig.~\ref{fig:pinn}A), where $\bm{\theta}_u$ is the set of trainable parameters in the network. The network takes the coordinates $\mathbf{x}$ as the input and outputs the approximate solution $\hat{\mathbf{u}}(\mathbf{x})$. Similarly, we also employ another, independent, fully connected network $\hat{\gamma}(\mathbf{x};\bm{\theta}_\gamma)$ for the unknown parameters $\gamma$ (Fig.~\ref{fig:pinn}A). We then restrict the two networks of $\hat{\mathbf{u}}$ and $\hat{\gamma}$ to satisfy the PDEs by using a PDE-informed loss function (Fig.~\ref{fig:pinn}A):
\begin{equation} \label{eq:loss_pde}
    \mathcal{L}_{\mathcal{F}}(\bm{\theta}_u,\bm{\theta}_\gamma) = \frac{1}{MN} \sum_{j=1}^M \sum_{i=1}^N \left| \mathcal{F}_i \left[ \hat{\mathbf{u}} (\mathbf{x}_j);\hat{\gamma}(\mathbf{x}_j) \right] \right|^2,
\end{equation}
where $\{\mathbf{x}_1, \mathbf{x}_2, \cdots, \mathbf{x}_M\}$ are a set of $M$ residual points in the domain $\Omega$, and $\left| \mathcal{F}_i \left[ \hat{\mathbf{u}}(\mathbf{x}_j); \hat{\gamma}(\mathbf{x}_j) \right] \right|$ measures the discrepancy of the $i$-th PDE $\mathcal{F}_i[\mathbf{u};\gamma]=0$ at the residual point $\mathbf{x}_j$.

There are multiple ways to sample the residual points, e.g., uniformly distributed random points or grid points, and in this study, we use a Sobol sequence to sample the residual points~\cite{pang2019fpinns}. $\mathcal{F}_i$ requires the derivatives of the network output $\hat{\mathbf{u}}$ with respect to the input $\mathbf{x}$ (e.g., $\nabla \hat{\mathbf{u}}$), which are evaluated exactly and efficiently via automatic differentiation (AD; also called ``backpropagation'' in deep learning) without generating a mesh. In order to compute arbitrary-order derivatives, we need to use a smooth activation function; in this study, we choose the hyperbolic tangent ($\tanh$). For more details of PINNs, we refer the reader to our review article~\cite{lu2019deepxde}.

\subsection{Hard-constraint boundary conditions} \label{sec:bc}

We can also enforce the BCs in Eq.~\eqref{eq:bc} via loss functions in the same way as the PDE loss in Eq.~\eqref{eq:loss_pde}. This approach can be used for all types of BCs, including Dirichlet, Neumann, Robin, or periodic BCs. In the examples of this study, we mainly consider Dirichlet and periodic BCs, and here we introduce another way to strictly impose the Dirichlet BCs and periodic BCs by modifying the network architecture. Compared to the approach of loss functions, this approach satisfies the BCs exactly and thus reduce the computational cost, and is also easier to be implemented.

\paragraph{Dirichlet BCs} Let us consider a Dirichlet BC for the solution $u_i$ ($1 \le i \le n$):
\begin{equation*}
    u_i(\mathbf{x}) = g(\mathbf{x}), \quad \mathbf{x} \in \Gamma_D,
\end{equation*}
where $\Gamma_D \subset \partial\Omega$ is a subset of the boundary. To make the approximate solution $\hat{u}_i(\mathbf{x};\bm{\theta}_u)$ satisfy this BC, we construct the solution as (Fig.~\ref{fig:pinn}B)
\begin{equation*}
    \hat{u}_i(\mathbf{x};\bm{\theta}_u) = g(\mathbf{x}) + \ell(\mathbf{x}) \mathcal{N}(\mathbf{x};\bm{\theta}_u),
\end{equation*}
where $\mathcal{N}(\mathbf{x};\bm{\theta}_u)$ is the network output, and $\ell$ is a function satisfying the following two conditions:
\begin{equation*}
    \left\{ \begin{array}{l}
    \ell(\mathbf{x}) = 0, \quad \mathbf{x} \in \Gamma_D, \\
    \ell(\mathbf{x}) > 0, \quad \mathbf{x} \in \Omega - \Gamma_D.
    \end{array}
    \right.
\end{equation*}
If $\Gamma_D$ is a simple geometry, then it is possible to choose $\ell(\mathbf{x})$ analytically~\cite{lagaris1998artificial,pang2019fpinns,lagari2020systematic}. For example, when $\Gamma_D$ is the boundary of an interval $\Omega = [a, b]$, i.e., $\Gamma_D = \{a, b\}$, we can choose $\ell(x)$ as $(x-a)(b-x)$ or $(1-e^{a-x})(1-e^{x-b})$. For complex domains, it is difficult to obtain an analytical formula of $\ell(\mathbf{x})$, and we can use spline functions to approximate $\ell(\mathbf{x})$~\cite{sheng2020pfnn}.

\paragraph{Periodic BCs} If $u_i(\mathbf{x})$ is a periodic function with respect to $x_j$ of the period $P$, then in the $x_j$ direction, $u_i(\mathbf{x})$ can be decomposed into a weighted summation of the basis functions of the Fourier series $\{ 1, \cos(\frac{2\pi x_j}{P}), \sin(\frac{2\pi x_j}{P}), \cos(\frac{4\pi x_j}{P}), \sin(\frac{4\pi x_j}{P}), \cdots \}$. Hence, we can replace the network input $x_j$ with the Fourier basis functions to impose the periodicity in the $x_j$ direction (Fig.~\ref{fig:pinn}C):
\begin{equation*}
    u_i(\mathbf{x}) = \mathcal{N}\left(x_1, \cdots, x_{j-1}, \left[ \cos(\frac{2\pi x_j}{P}), \sin(\frac{2\pi x_j}{P}), \cos(\frac{4\pi x_j}{P}), \sin(\frac{4\pi x_j}{P}),\cdots \right], x_{j+1}, \cdots, x_d\right).
\end{equation*}
In classical Fourier analysis, many basis functions may be required approximate an arbitrary periodic function with a good accuracy, but as demonstrated in~\cite{zhang2020learning}, we can use as few as two terms $\{ \cos(\frac{2\pi x_j}{P}), \sin(\frac{2\pi x_j}{P})\}$ without loss of accuracy, because all the other basis functions $\{ \cos(\frac{4\pi x_j}{P}), \sin(\frac{4\pi x_j}{P}), \cdots \}$ can be written as a nonlinear continuous function of $\cos(\frac{2\pi x_j}{P})$ and $\sin(\frac{2\pi x_j}{P})$ and neural networks are universal approximators of nonlinear continuous functions. Here, we consider the case where $u_i$ and all its derivatives are periodic; this approach can also be extended to the case where its derivatives up to a finite order is periodic~\cite{dong2020method}.

\subsection{Soft constraints} \label{sec:soft}

While the BCs in Eq.~\eqref{eq:constraint} can be imposed directly during constrained optimization, it is difficult to satisfy the PDEs and inequality constraint exactly. The simplest way to deal with these constraints is to consider them as soft constraints via loss functions. Specifically, using the PDE loss in Eq.~\eqref{eq:loss_pde}, we convert the original constrained optimization to an unconstrained optimization problem:
\begin{equation} \label{eq:soft}
    \min_{\bm{\theta}_u,\bm{\theta}_\gamma} \mathcal{L}(\bm{\theta}_u,\bm{\theta}_\gamma) =  \mathcal{J} + \mu_\mathcal{F}\mathcal{L}_{\mathcal{F}} + \mu_h\mathcal{L}_h,
\end{equation}
where $\mathcal{L}_h$ is a quadratic penalty to measure the violation of the hard constraint $h(\mathbf{u}, \gamma) \le 0$:
\begin{equation} \label{eq:loss_h}
    \mathcal{L}_h(\bm{\theta}_u,\bm{\theta}_\gamma) = \mathbbm{1}_{\{h(\hat{\mathbf{u}}, \hat{\gamma}) > 0\}} h^2(\hat{\mathbf{u}}, \hat{\gamma}),
\end{equation}
and $\mu_\mathcal{F}$ and $\mu_h$ are the fixed penalty coefficients of the soft constraints. Then the final solution is obtained by minimizing the total loss via gradient-based optimizers:
\begin{equation*}
    \bm{\theta}^*_u,\bm{\theta}^*_\gamma = \arg\min_{\bm{\theta}_u,\bm{\theta}_\gamma} \mathcal{L}(\bm{\theta}_u,\bm{\theta}_\gamma).
\end{equation*}

If $\mu_\mathcal{F}$ and $\mu_h$ are larger, we penalize the constraint violations more severely, thereby forcing the solution satisfying the constraints better. However, when the penalty coefficients are too large, the optimization problem becomes ill-conditioned and hence makes it difficult to converge to a minimum~\cite{bertsekas2014constrained,nocedal2006numerical}. On the other hand, if the penalty coefficients are too small, then the obtained solution will not satisfy the constraints and thus is not a valid solution. Therefore, although this approach is simple, it cannot be used in general. In contrast, the soft-constraint approach has worked well for inverse problems to match observed measurements~\cite{chen2020physics,raissi2020hidden,yazdani2020systems}, as we discussed in Section~\ref{sec:intro}.

\subsection{Penalty method} \label{sec:penalty}

To overcome the optimization difficulty of the soft constraints with large penalty coefficients, we consider the penalty method. Unlike the approach of soft constraints, which converts a constrained optimization problem to an unconstrained optimization problem with fixed coefficients, a penalty method replaces the constrained optimization problem by a \emph{sequence} of unconstrained problems with varying coefficients. The unconstrained problem in the $k$-th ``outer'' iteration is
\begin{equation*}
    \min_{\bm{\theta}_u,\bm{\theta}_\gamma} \mathcal{L}^k(\bm{\theta}_u,\bm{\theta}_\gamma) =  \mathcal{J} + \mu_\mathcal{F}^k\mathcal{L}_{\mathcal{F}} + \mu_h^k\mathcal{L}_h,
\end{equation*}
where $\mu_\mathcal{F}^k$ and $\mu_h^k$ are the penalty coefficients in the $k$-th iteration. In each iteration, we increase the penalty coefficients by constant factors $\beta_\mathcal{F} > 1$ and $\beta_h>1$:
\begin{equation*}
    \mu_\mathcal{F}^{k+1}=\beta_\mathcal{F}\mu_\mathcal{F}^k, \quad \mu_h^{k+1}=\beta_h\mu_h^k.
\end{equation*}
Here, we need to choose the initial coefficients $\mu_\mathcal{F}^0$ and $\mu_h^0$ and the factors $\beta_\mathcal{F}$ and $\beta_h$. The choice of these hyperparameters are problem dependent~\cite{bertsekas2014constrained}. Similar to the soft-constraint approach, we would have an ill-conditioned optimization problem when their values are large, leading to slow convergence. However, if their values are small, we may need many outer iterations, and gradient-descent optimization may get stuck at poor local minima, as we will show in our numerical experiments. Algorithm \ref{alg:penalty} presents the pseudocode for the penalty method.

\begin{algorithm}[htbp]
\caption{hPINNs via the penalty method.}
\label{alg:penalty}
\begin{algorithmic}
\STATE \textbf{Hyperparameters}: initial penalty coefficients $\mu_\mathcal{F}^0$ and $\mu_h^0$, factors $\beta_\mathcal{F}$ and $\beta_h$
\STATE $k \longleftarrow 0$
\STATE $\bm{\theta}_u^0,\bm{\theta}_\gamma^0 \longleftarrow \arg\min_{\bm{\theta}_u,\bm{\theta}_\gamma} \mathcal{L}^0(\bm{\theta}_u,\bm{\theta}_\gamma)$: Train the networks $\hat{\mathbf{u}}(\mathbf{x};\bm{\theta}_u)$ and $\hat{\gamma}(\mathbf{x};\bm{\theta}_\gamma)$ from random initialization, until the training loss is converged
\REPEAT
    \STATE $k \longleftarrow k + 1$
    \STATE $\mu_\mathcal{F}^{k} \longleftarrow \beta_\mathcal{F}\mu_\mathcal{F}^{k-1}$
    \STATE $\mu_h^{k} \longleftarrow \beta_h\mu_h^{k-1}$
    \STATE $\bm{\theta}_u^k,\bm{\theta}_\gamma^k \longleftarrow \arg\min_{\bm{\theta}_u,\bm{\theta}_\gamma} \mathcal{L}^k(\bm{\theta}_u,\bm{\theta}_\gamma)$: Train the networks $\hat{\mathbf{u}}(\mathbf{x};\bm{\theta}_u)$ and $\hat{\gamma}(\mathbf{x};\bm{\theta}_\gamma)$ from the initialization of $\bm{\theta}_u^{k-1}$ and $\bm{\theta}_\gamma^{k-1}$, until the training loss is converged
\UNTIL{$\mathcal{L}_\mathcal{F}(\bm{\theta}_u^k,\bm{\theta}_\gamma^k)$ and $\mathcal{L}_h(\bm{\theta}_u^k,\bm{\theta}_\gamma^k)$ are smaller than a tolerance}
\end{algorithmic}
\end{algorithm}

As $k \to \infty$, given that the networks are well trained, the solutions of the successive unconstrained optimization problems will converge to the solution of the original constrained optimization problem~\cite{bertsekas2014constrained,nocedal2006numerical}. In the first iteration ($k=0$), the neural networks are trained from a random initialization, while the neural networks in the ($k+1$)-th iteration are trained by using the solution of the $k$-th iteration as the initialization---this good starting point helps to counteract the slow convergence that would otherwise arise when $\mu_\mathcal{F}^k$ and $\mu_h^k$ are large.

\subsection{Augmented Lagrangian method}

The third method we considered in this study is the method of multipliers or the augmented Lagrangian method~\cite{toussaint2014introduction}. Similar to penalty methods, the augmented Lagrangian method also uses penalty terms, but it adds new terms designed to mimic Lagrange multipliers. The unconstrained problem in the $k$-th iteration is
\begin{equation} \label{eq:auglag}
    \begin{array}{rl}
    \min_{\bm{\theta}_u,\bm{\theta}_\gamma} \mathcal{L}^k(\bm{\theta}_u,\bm{\theta}_\gamma) =&  \mathcal{J} \\
    &+\, \mu_\mathcal{F}^k\mathcal{L}_{\mathcal{F}} \\
    &+\, \mu_h^k \mathbbm{1}_{\{h> 0\, \lor\, \lambda_h^k > 0\}} h^2 \\
    &+\, \frac{1}{MN}\sum_{j=1}^M\sum_{i=1}^N \lambda_{i,j}^k \mathcal{F}_i \left[ \hat{\mathbf{u}} (\mathbf{x}_j);\hat{\gamma}(\mathbf{x}_j) \right] \\
    &+\, \lambda_h^k h,
    \end{array}
\end{equation}
where the symbol ``$\lor$'' in the third term is the logical OR, and $\lambda_{i,j}^k$ and $\lambda_h^k$ are multipliers. We note that the second penalty term $\mathcal{L}_\mathcal{F}$ is the same as the term in Eq.~\eqref{eq:loss_pde} in the soft constraints and penalty method, but the third penalty term of $h$ depends on the multiplier $\lambda_h^k$, and is slightly different from the penalty term $\mathcal{L}_h$ in Eq.~\eqref{eq:loss_h}.

The last two terms in Eq.~\eqref{eq:auglag} are Lagrangian terms, and we take the term $\mathcal{F}_i \left[ \hat{\mathbf{u}} (\mathbf{x}_j);\hat{\gamma}(\mathbf{x}_j) \right]$ as an example to demonstrate its effect and how to choose $\lambda_{i,j}^k$. It is clear that the gradient $\nabla \mathcal{F}_i \left[ \hat{\mathbf{u}} (\mathbf{x}_j);\hat{\gamma}(\mathbf{x}_j)\right]$ is always orthogonal to the constraint of $\mathcal{F}_i \left[ \hat{\mathbf{u}} (\mathbf{x}_j);\hat{\gamma}(\mathbf{x}_j) \right]$. In the $k$-th iteration, we choose $\lambda_{i,j}^k$ to generate exactly the gradient that was previously generated in the $(k-1)$-th iteration by the penalty term $\left| \mathcal{F}_i \left[ \hat{\mathbf{u}} (\mathbf{x}_j);\hat{\gamma}(\mathbf{x}_j) \right] \right|^2$ in $\mathcal{L}_\mathcal{F}$~\cite{bertsekas2014constrained,nocedal2006numerical,toussaint2014introduction}, i.e., we require that
\begin{multline*}
    \lambda_{i,j}^k \nabla \mathcal{F}_i \left[ \hat{\mathbf{u}} (\mathbf{x}_j;\bm{\theta}_u^{k-1}); \hat{\gamma}(\mathbf{x}_j;\bm{\theta}_\gamma^{k-1})\right] = \\
    \mu_\mathcal{F}^{k-1} \nabla \left| \mathcal{F}_i \left[ \hat{\mathbf{u}} (\mathbf{x}_j;\bm{\theta}_u^{k-1});\hat{\gamma}(\mathbf{x}_j;\bm{\theta}_\gamma^{k-1}) \right] \right|^2 + \lambda_{i,j}^{k-1} \nabla \mathcal{F}_i \left[ \hat{\mathbf{u}} (\mathbf{x}_j;\bm{\theta}_u^{k-1}); \hat{\gamma}(\mathbf{x}_j;\bm{\theta}_\gamma^{k-1})\right],
\end{multline*}
and thus we have
\begin{equation*}
    \lambda_{i,j}^k = \lambda_{i,j}^{k-1} + 2\mu_\mathcal{F}^{k-1} \mathcal{F}_i \left[ \hat{\mathbf{u}} (\mathbf{x}_j;\bm{\theta}_u^{k-1});\hat{\gamma}(\mathbf{x}_j;\bm{\theta}_\gamma^{k-1}) \right].
\end{equation*}
Similarly, for $\lambda_h^k$, we have~\cite{toussaint2014introduction}
\begin{equation*}
    \lambda_h^k = \max \left(\lambda_h^{k-1} + 2\mu_h^{k-1}h \left(\hat{\mathbf{u}}(\mathbf{x};\bm{\theta}_u^{k-1}), \hat{\gamma}(\mathbf{x};\bm{\theta}_\gamma^{k-1}) \right), 0 \right).
\end{equation*}
The initial values of all the multipliers are chosen as 0. We will show in our numerical examples that $\lambda_{i,j}^k/\mu_\mathcal{F}^{k}$ and $\lambda_h^k/\mu_h^{k}$ converge after several iterations. The pseudocode is presented in Algorithm \ref{alg:auglag}. Compared to the penalty method, the augmented Lagrangian method has two main advantages: (1) it is not necessary to increase $\mu_\mathcal{F}$ and $\mu_h$ to infinity in order to induce convergence to a feasible solution, which further avoids the ill-conditioning; (2) the convergence rate is considerably better than that of the penalty method.

\begin{algorithm}[htbp]
\caption{hPINNs via the augmented Lagrangian method.}
\label{alg:auglag}
\begin{algorithmic}
\STATE \textbf{Hyperparameters}: initial penalty coefficients $\mu_\mathcal{F}^0$ and $\mu_h^0$, factors $\beta_\mathcal{F}$ and $\beta_h$
\STATE $k \longleftarrow 0$
\STATE $\lambda_{i,j}^0 \longleftarrow 0$ for $1 \le i \le N,\, 1 \le j \le M$
\STATE $\lambda_h^0 \longleftarrow 0$
\STATE $\bm{\theta}_u^0,\bm{\theta}_\gamma^0 \longleftarrow \arg\min_{\bm{\theta}_u,\bm{\theta}_\gamma} \mathcal{L}^0(\bm{\theta}_u,\bm{\theta}_\gamma)$: Train the networks $\hat{\mathbf{u}}(\mathbf{x};\bm{\theta}_u)$ and $\hat{\gamma}(\mathbf{x};\bm{\theta}_\gamma)$ from random initialization, until the training loss is converged
\REPEAT
    \STATE $k \longleftarrow k + 1$
    \STATE $\mu_\mathcal{F}^{k} \longleftarrow \beta_\mathcal{F}\mu_\mathcal{F}^{k-1}$
    \STATE $\mu_h^{k} \longleftarrow \beta_h\mu_h^{k-1}$
    \STATE $\lambda_{i,j}^k \longleftarrow \lambda_{i,j}^{k-1} + 2\mu_\mathcal{F}^{k-1} \mathcal{F}_i \left[ \hat{\mathbf{u}} (\mathbf{x}_j;\bm{\theta}_u^{k-1});\hat{\gamma}(\mathbf{x}_j;\bm{\theta}_\gamma^{k-1}) \right]$ for $1\le i\le N,\, 1\le j\le M$
    \STATE $\lambda_h^k \longleftarrow \max \left(\lambda_h^{k-1} + 2\mu_h^{k-1}h \left(\hat{\mathbf{u}}(\mathbf{x};\bm{\theta}_u^{k-1}), \hat{\gamma}(\mathbf{x};\bm{\theta}_\gamma^{k-1}) \right), 0 \right)$
    \STATE $\bm{\theta}_u^k,\bm{\theta}_\gamma^k \longleftarrow \arg\min_{\bm{\theta}_u,\bm{\theta}_\gamma} \mathcal{L}^k(\bm{\theta}_u,\bm{\theta}_\gamma)$: Train the networks $\hat{\mathbf{u}}(\mathbf{x};\bm{\theta}_u)$ and $\hat{\gamma}(\mathbf{x};\bm{\theta}_\gamma)$ from the initialization of $\bm{\theta}_u^{k-1}$ and $\bm{\theta}_\gamma^{k-1}$, until the training loss is converged
\UNTIL{$\mathcal{L}_\mathcal{F}(\bm{\theta}_u^k,\bm{\theta}_\gamma^k)$ and $\mathcal{L}_h(\bm{\theta}_u^k,\bm{\theta}_\gamma^k)$ are smaller than a tolerance}
\end{algorithmic}
\end{algorithm}

\section{Results} \label{sec:res}

We will apply our proposed hPINNs to solve two different problems of inverse design in optics and fluids. We compare hPINNs with a few traditional PDE-constrained optimization methods based on the adjoint method and a numerical PDE solver, and demonstrate the capacity and effectiveness of hPINNs. All the hPINN codes in this study are implemented by using the library DeepXDE~\cite{lu2019deepxde}, and 
will be
% have been
deposited in GitHub at \url{https://github.com/lululxvi/hpinn}.

\subsection{Holography}

We first consider the challenging problem of holography for an image in depth (in the direction of propagation). In contrast to holograms of images that are parallel to the scattering surface, and which can be designed via algorithms relying on Fourier optics~\cite{goodman2005introduction}, in-depth holograms require us to use inverse design on the full Maxwell’s equations. We propose to design the permittivity map of a scattering slab, that scatters light so that the transmitted intensity has a targeted shape.

\subsubsection{Problem setup}

\begin{figure}[htbp]
    \centering
    \includegraphics{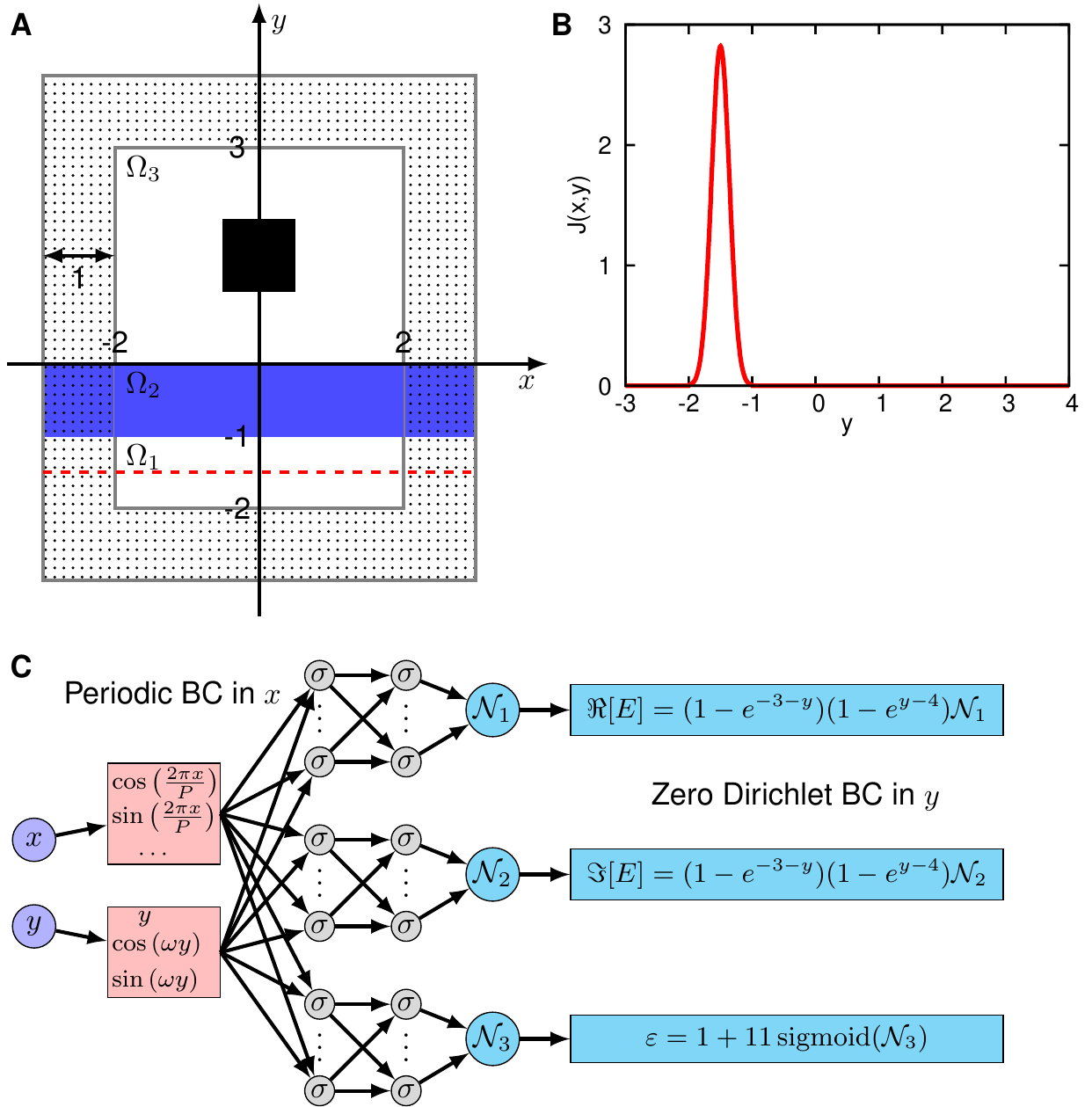}
    \caption{\textbf{Holography problem setup and the neural network architecture.} (\textbf{A}) The whole computational domain includes the main domain $\Omega = [-2,-2] \times [2,3]$ and a PML of depth one in the shaded region. The design region for the permittivity $\varepsilon$ is in blue, and the center of the current $J$ is the dashed red line in the domain $\Omega_1$. The target electrical field is defined in $\Omega_3$. (\textbf{B}) The time-harmonic current $J$. (\textbf{C}) The architecture of the hPINN with the Dirichlet and periodic BCs embedded directly in the network. The network inputs are $x$ and $y$, and the outputs are $\Re[E]$, $\Im[E]$ and $\varepsilon$.}
    \label{fig:holography}
\end{figure}

We consider a holography problem defined on a rectangle domain $\Omega = [-2,-2] \times [2,3]$ (Fig.~\ref{fig:holography}A). In the lower part of $\Omega$ (i.e., $\Omega_1$), we have a time-harmonic current $J$ to generate an electromagnetic wave $E(x,y)=\Re[E] + i\Im[E]$ in the whole space. There is a lens in the region $\Omega_2$ (the blue region), whose permittivity function $\varepsilon(x,y)$ (the square of the refractive index) is to be designed to produce our target transmitted-wave pattern $f(x,y)$ in the top region $\Omega_3 = [-2, 0] \times [2, 3]$, while the permittivity in other region is one. Specifically, our objective function in this problem is
\begin{equation} \label{eq:holography_J}
    \mathcal{J}(E) = \frac{1}{\text{Area}(\Omega_3)} \| |E(x,y)|^2 - f(x,y) \|^2_{2,\Omega_3} = \frac{1}{\text{Area}(\Omega_3)} \int_{\Omega_3} \left(|E(x,y)|^2 - f(x,y)\right)^2 dxdy,
\end{equation}
where $|E|^2 = (\Re[E])^2 + (\Im[E])^2$ is the square of the magnitude of the electric field. In this study, we choose the target function as
\begin{equation*}
    f(x,y) = \left\{ \begin{array}{ll}
    1, & (x,y) \in [-0.5,0.5] \times [1,2] \\
    0, & \text{otherwise}
    \end{array} \right. ,
\end{equation*}
i.e., $f$ is equal to 1 inside the black square in Fig.~\ref{fig:holography}A and 0 otherwise. 
We can think of the target as an input defined by a user, who does not need to know Helmholtz's equation. Still in that case, our inverse design tool should find the geometry that scatters the transmitted intensity in a shape as close to the user input as possible. In the case of our example, the target function is not a solution of Helmholtz's equation (equivalent to Maxwell's equations in two dimensions). So, we expect the neural network to have to make a trade-off between reaching the target shape and making sure that Helmholtz's equations are accurately solved.

The holography problem can be described by the following PDEs:
\begin{equation} \label{eq:wave}
    \nabla^2E + \varepsilon\omega^2E =-i\omega J,
\end{equation}
where the frequency $\omega$ is chosen as $2\pi$ (corresponding to a wavelength $1$ in the $\varepsilon=1$ region), and the electric current source $J$ is chosen as a Gaussian profile in $y$ (Fig.~\ref{fig:holography}B) and a constant in $x$, which generates an incident planewave propagating in the $y$ direction:
\begin{equation*}
    J(x,y)=\frac{1}{h\sqrt{\pi}}e^{-(\frac{y+1.5}{h})^2} \mathbbm{1}_{[-1,-2]}(y) ,
\end{equation*}
where $h=0.2$ and $\mathbbm{1}_{[-1,-2]}$ is the indicator function that truncates $J$ to be supported in a finite-width strip $y \in [-1,-2]$ (centered on the dashed-red line in Fig.~\ref{fig:holography}A). This problem is formally defined in an infinitely large domain with outgoing (radiation) boundary conditions. To reduce the problem to a finite domain for computation, we apply the technique of perfectly matched layers (PMLs)~\cite{johnson2008notes}, which works as an artificial absorbing layer (the black-dotted region in Fig.~\ref{fig:holography}A) to truncate the computational domain. After applying the PML, the PDE in Eq.~\eqref{eq:wave} becomes
\begin{equation} \label{eq:pml}
\frac{1}{1+i\frac{\sigma_x(x)}{\omega}} \frac{\partial}{\partial x} \left(\frac{1}{1+i\frac{\sigma_x(x)}{\omega}} \frac{\partial E}{\partial x}\right) + \frac{1}{1+i\frac{\sigma_y(y)}{\omega}}\frac{\partial}{\partial y} \left(\frac{1}{1+i\frac{\sigma_y(y)}{\omega}} \frac{\partial E}{\partial y}\right) + \varepsilon\omega^2E =-i\omega J,
\end{equation}
where
$$\sigma_x(x)=\sigma_0 (-2-x)^2 \mathbbm{1}_{(-\infty, -2)}(x) + \sigma_0 (x-2)^2 \mathbbm{1}_{(2,\infty)}(x),$$
$$\sigma_y(y)= \sigma_0 (-2-y)^2 \mathbbm{1}_{(-\infty, -2)}(y) +\sigma_0 (y-3)^2 \mathbbm{1}_{(3,\infty)}(y),$$
with $\sigma_0=-\ln 10^{-20} / (4 d^3/3) \gg 1$ and $d=1$ is the depth of the PML layer. For BCs of the PML, we use a periodic BC for the $x$ direction ($x=-3$ and $x=3$) and a zero Dirichlet BC for the $y$ direction ($y=-3$ and $y=4$). Corresponding to semiconductor dielectric materials commonly used infrared optics, we require $\varepsilon \in [1,12]$ in the design region.

\subsubsection{hPINN}

We will apply hPINN to solve this inverse design problem of holography. The objective function is defined in Eq.~\eqref{eq:holography_J} and is approximated by Monte-Carlo integration. The PDEs are the real and imaginary parts of Eq.~\eqref{eq:pml} (see Section \ref{sec:holography_pde}), i.e., $N=2$, and the PDE-informed loss function is
\begin{equation} \label{eq:holography_loss_pde}
    \mathcal{L}_{\mathcal{F}} = \frac{1}{2M} \sum_{j=1}^M \left( \Re[\mathcal{F} \left[ \mathbf{x}_j \right] ] \right)^2 + \left( \Im[\mathcal{F} \left[ \mathbf{x}_j \right] ] \right)^2,
\end{equation}
where
\begin{equation*}
    \mathcal{F}[\mathbf{x}_j] = \frac{1}{\omega +i\sigma_x(x)} \frac{\partial}{\partial x} \left(\frac{1}{1+i\frac{\sigma_x(x)}{\omega}} \frac{\partial E}{\partial x}\right) + \frac{1}{\omega+i\sigma_y(y)}\frac{\partial}{\partial y} \left(\frac{1}{1+i\frac{\sigma_y(y)}{\omega}} \frac{\partial E}{\partial y}\right) + \varepsilon\omega E + i J \Bigg|_{\mathbf{x}_j}.
\end{equation*}
Here, we scaled the PDE in Eq.~\eqref{eq:pml} by $\frac{1}{\omega}$ to make $\mathcal{F}$ of order one. In this problem, we do not have any inequality constraint. 

We construct three networks to approximate $\Re[E]$, $\Im[E]$, and $\varepsilon$ (Fig.~\ref{fig:holography}C), respectively. The periodic and Dirichlet BCs are imposed directly into the network. The restriction of $\varepsilon \in [1,12]$ is satisfied by using the transformation $\varepsilon(x,y) = 1+11\, \text{sigmoid}(\mathcal{N}_3(x,y))$, where $\text{sigmoid}(x) = \frac{1}{1+e^{-x}}$ and $\mathcal{N}_3(x,y)$ is a network output. In addition, as demonstrated in~\cite{yazdani2020systems}, it is usually beneficial to the network training by adding extra features, which may have a similar pattern of the solution (even if they are not accurate), to the network input. In our study, we use the two extra features $\cos(\omega y)$ and $\sin(\omega y)$, because we expect the current $J$ to generate an incident plane wave with the frequency $\omega$ in the $y$ direction (in addition to scattered waves from $\varepsilon \ne 1$ regions).

\subsubsection{Hyperparameters and verification on a forward problem}

To verify that hPINN is capable of solving the holography problem, we first solve a forward problem, where $\varepsilon=1$ is given and we only optimize the network to minimize the loss $\mathcal{L}_\mathcal{F}$ in Eq.~\eqref{eq:holography_loss_pde}. We choose $M=17000$ such that the average spacing between two adjacent random points is $\sim 0.05$. Each network in Fig.~\ref{fig:holography}C has 5 layers with 32 neurons per layer, and we use 4 Fourier basis $\{\cos\left(\frac{2\pi x}{P}\right), \sin\left(\frac{2\pi x}{P}\right), \cos\left(\frac{4\pi x}{P}\right), \sin\left(\frac{4\pi x}{P}\right)\}$. To train the networks, we first use Adam optimizer~\cite{kingma2014adam} with the learning rate \num{1e-3} for \num{2e4} steps, and then switch to L-BFGS~\cite{byrd1995limited} until the loss is converged.

\begin{figure}[htbp]
    \centering
    \includegraphics[width=\textwidth]{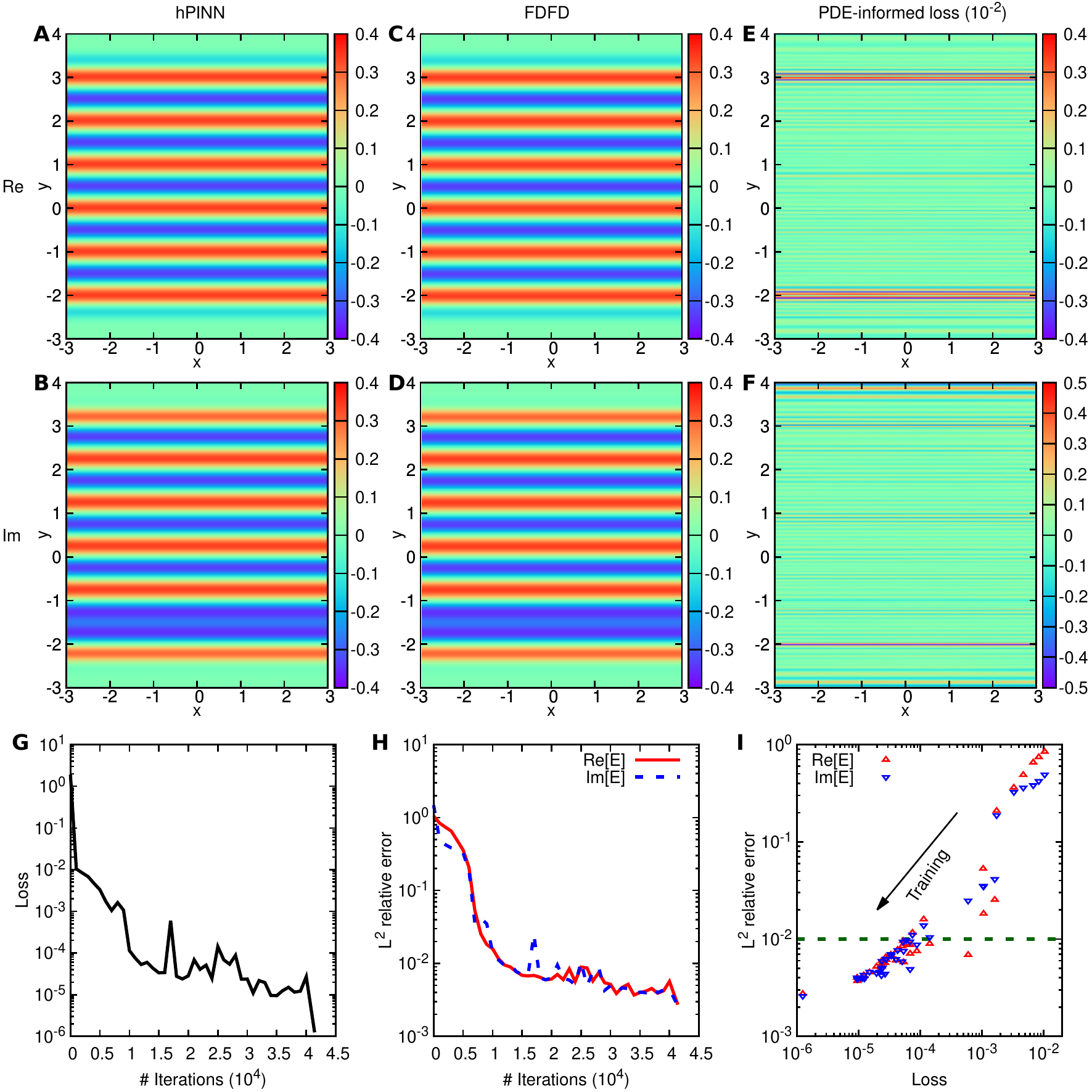}
    \caption{\textbf{hPINN for solving a forward problem.} (\textbf{A} and \textbf{B}) The (A) real part $\Re[E]$ and (B) imaginary part $\Im[E]$ of the hPINN solution. (\textbf{C} and \textbf{D}) The (C) real part $\Re[E]$ and (D) imaginary part $\Im[E]$ of the FDFD solution. (\textbf{E} and \textbf{F}) The (E) real part $\Re[\mathcal{F}[\mathbf{x}]]$ and (F) imaginary part $\Im[\mathcal{F}[\mathbf{x}]]$ of the PDE-informed loss in Eq.~\eqref{eq:holography_loss_pde}. (\textbf{G} and \textbf{H}) The (G) training loss $\mathcal{L}_\mathcal{F}$ and (H) $L^2$ relative error versus the number of optimization iterations. (\textbf{I}) The correlation between the training loss and the $L^2$ relative error during the training process.}
    \label{fig:holography_forward}
\end{figure}

After the network training, the solution of hPINN (Figs. \ref{fig:holography_forward}A and B) is consistent with the reference solution obtained from the finite-difference frequency-domain (FDFD) method~\cite{champagne2001fdfd} with the spatial resolutions $\Delta x = 0.01$ (Figs. \ref{fig:holography_forward}C and D). We also show that the pointwise PDE-informed loss $\mathcal{F}[\mathbf{x}]$ is very small (between \num{-5e-3} and \num{5e-3}; Figs. \ref{fig:holography_forward}E and F). During the training process, the loss function decreases (Fig.~\ref{fig:holography_forward}G), while the $L^2$ relative error computed using the reference solution is also decreases (Fig.~\ref{fig:holography_forward}H). There is a clear correlation between the training loss and the $L^2$ relative error (Fig.~\ref{fig:holography_forward}I), and when the training loss is $\lesssim 10^{-4}$, the $L^2$ relative error is $< 1\%$. This criterion is useful for the following reverse design problem, because we can easily check the accuracy of the hPINN solution during the training process by monitoring the loss $\mathcal{L}_\mathcal{F}$ directly without comparing to the FDFD reference.

\subsubsection{Soft constraints}

\begin{figure}[htbp]
    \centering
    \includegraphics[width=\textwidth]{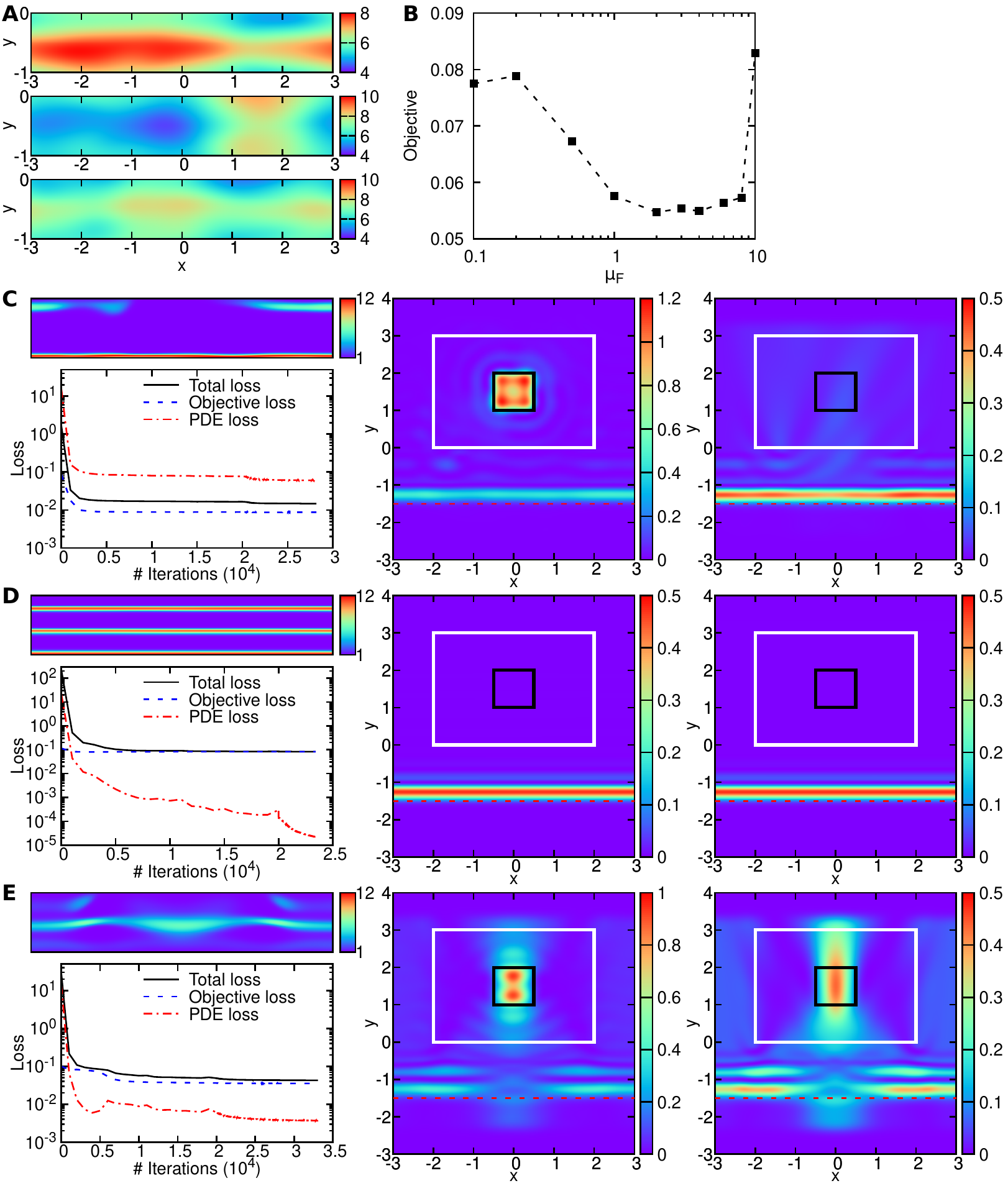}
    \caption{\textbf{hPINN for the inverse design of holography via the approach of soft constraints.} (\textbf{A}) Three examples of  random initialization of $\varepsilon$. (\textbf{B}) The objective $\mathcal{J}$ of different designs obtained from hPINNs with different $\mu_\mathcal{F}$. (\textbf{C}, \textbf{D}, and \textbf{E}) The (left top) permittivity $\varepsilon$, (left bottom) training trajectory, (center) hPINN solution $|E|^2$, and (right) the reference $|E|^2$ obtained from FDFD for (C) $\mu_\mathcal{F}=0.1$, (D) $\mu_\mathcal{F}=10$, and (E) $\mu_\mathcal{F}=2$.}
    \label{fig:holography_soft}
\end{figure}

We first use the approach of soft constraints to solve the holography inverse-design problem by minimizing
\begin{equation*}
    \mathcal{L} = \mathcal{J} + \mu_\mathcal{F} \mathcal{L}_\mathcal{F}.
\end{equation*}
The permittivity function $\varepsilon$ is randomly initialized, and three random examples are shown in Fig.~\ref{fig:holography_soft}A. The objective value for a random permittivity function is $\sim 0.1$. As we discussed in Section \ref{sec:soft}, the choice of $\mu_\mathcal{F}$ plays an important role in the final design, and we will compare the performance of different values of $\mu_\mathcal{F}$ using the first case in Fig.~\ref{fig:holography_soft}A as the initial permittivity.

We use the same hyperparameters and training procedure as we used in the forward problem. After the network training, the third network is the permittivity function, but we do not use the electric field (the first and second networks) to validate the result and determine the actual performance, because we will show that the electric field of the network is not accurate. Instead, we use FDFD to simulate the correct electric field from the permittivity function, and then compute the objective $\mathcal{J}$. We show that when $\mu_\mathcal{F}$ is too large or too small, the final objective is relatively large, and the smallest objective is obtained when $\mu_\mathcal{F} \approx 2$ (Fig.~\ref{fig:holography_soft}B).

When $\mu_\mathcal{F}$ is small, e.g., 0.1, the PDE loss cannot be optimized well ($\mathcal{L}_\mathcal{F} \sim 10^{-1}$; Fig.~\ref{fig:holography_soft}C left bottom), and thus the PDEs are not satisfied well. If we use the obtained permittivity function (Fig.~\ref{fig:holography_soft}C left top) to simulate the corresponding electric field $|E|^2$ via FDFD (Fig.~\ref{fig:holography_soft}C right), this is very different from the $|E|^2$ obtained from hPINN (Fig.~\ref{fig:holography_soft}C center). On the other hand, when $\mu_\mathcal{F}$ is large, e.g., 10, the final PDE loss is $\sim 10^{-5}$, and the field of $|E|^2$ from hPINN and FDFD are almost identical (Fig.~\ref{fig:holography_soft}D), i.e., the PDE constraints are satisfied very well. However, the permittivity function is not meaningful and the objective is very large (Fig.~\ref{fig:holography_soft}D), because of the ill-conditioning of the optimization. The case of the smallest objective ($\approx 0.0547$) with $\mu_\mathcal{F}=2$ is shown in Fig.~\ref{fig:holography_soft}E, but the prediction of $|E|^2$ is still of low accuracy. Therefore, the soft constraint approach cannot satisfy the PDE constraints for small $\mu_\mathcal{F}$, and thus a large $\mu_\mathcal{F}$ is required. However, for large $\mu_\mathcal{F}$, the optimization failed to make progress on the design objective due to the ill-conditioning.

\subsubsection{Penalty method}

To address the issue in the soft constraint approach, we gradually increase the value of $\mu_\mathcal{F}$ by using the penalty method in Algorithm \ref{alg:penalty}. The initial value of $\mu_\mathcal{F}$ is $\mu_\mathcal{F}^0$ and the increasing factor is $\beta_\mathcal{F}$, i.e., $\mu_\mathcal{F}^k = (\beta_\mathcal{F})^k \mu_\mathcal{F}^0$.

\begin{figure}[htbp]
    \centering
    \includegraphics[width=\textwidth]{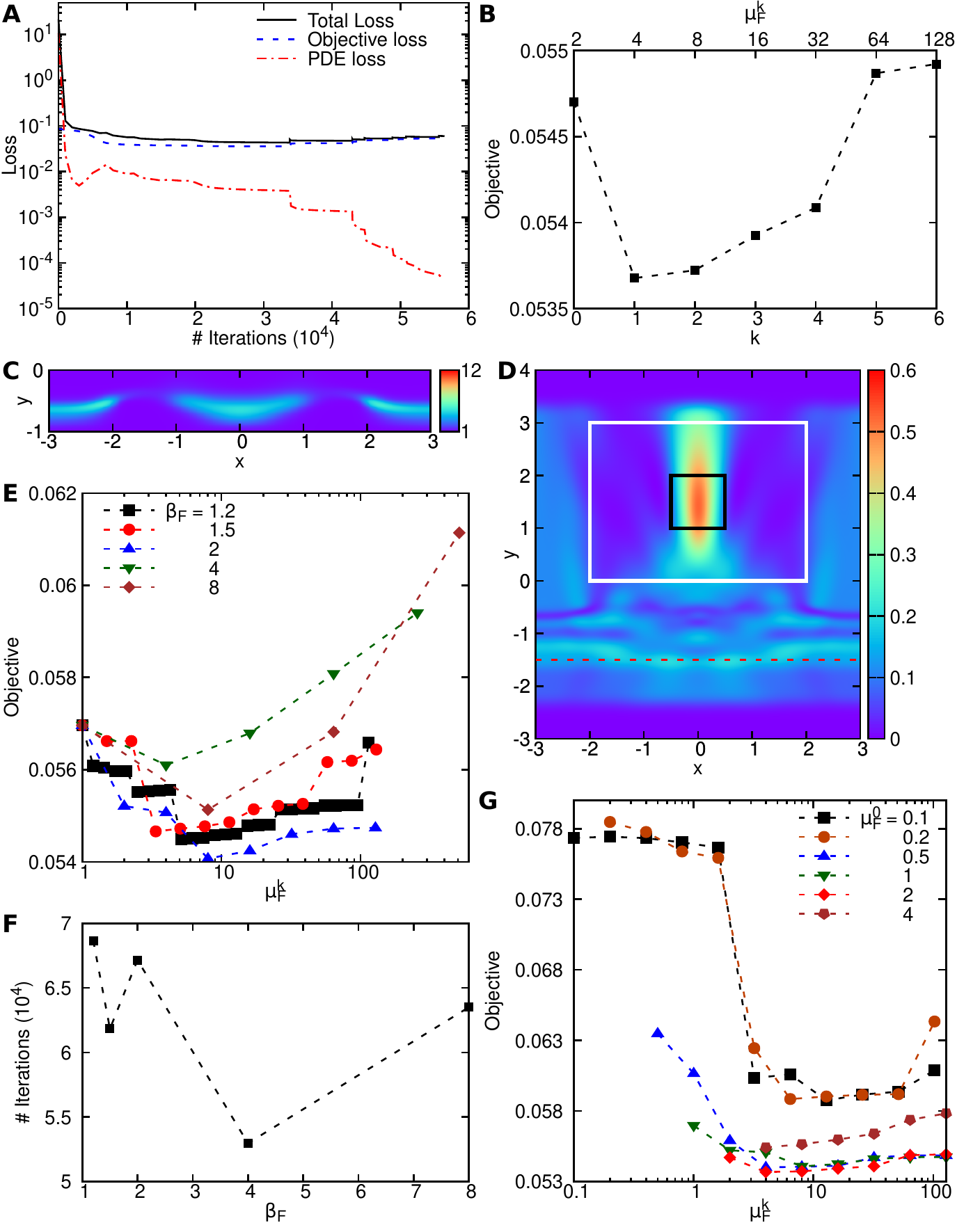}
    \caption{\textbf{hPINN for the inverse design of holography via the penalty method.} (\textbf{A} to \textbf{D}) The results of $\mu_\mathcal{F}^0=2$ and $\beta_\mathcal{F}=2$. (A) The losses versus the number of optimization iterations, (B) the objective value after the network is trained with $\mu_\mathcal{F}^k$, (C) the optimized permittivity function $\varepsilon$ at $k=1$, and (D) the corresponding electric field $|E|^2$. (\textbf{E}) The objective value versus $\mu_\mathcal{F}^k$ when using different values of $\beta_\mathcal{F}$. (\textbf{F}) The total number of optimization iterations (which is proportional to the computational cost) for different $\beta_\mathcal{F}$. (\textbf{G}) The objective value versus $\mu_\mathcal{F}^k$ for different values of $\mu_\mathcal{F}^0$..}
    \label{fig:holography_penalty}
\end{figure}

We need to tune $\mu_\mathcal{F}^0$ and $\beta_\mathcal{F}$ in the penalty method. As an example, we first show the results of the case $\mu_\mathcal{F}^0=2$ and $\beta_\mathcal{F}=2$ (Figs. \ref{fig:holography_penalty}A--D). When we increase the value of $\mu_\mathcal{F}$, the PDE loss decreases (Fig.~\ref{fig:holography_penalty}A), and we find that in order to make $\mathcal{L}_\mathcal{F} < 10^{-4}$, we need to increase $\mu_\mathcal{F}$ until $\mu_\mathcal{F} > 100$. After we train the network with $\mu_\mathcal{F}^k$, the objective first decreases and then increases (Fig.~\ref{fig:holography_penalty}B). In the soft-constraint approach, when $\mu_\mathcal{F}=10$, the network would get stuck at a poor local minimum due to the ill-conditioning, but here, even if $\mu_\mathcal{F} > 100$, we still obtain a meaningful result and objective. However, the objective still becomes worse when $\mu_\mathcal{F}^k > 10$, and thus the ill-conditioning problem still exists, albeit weaker than before. The smallest objective $\sim 0.0537$ is obtained at $k=1$, which is $\sim 2\%$ better than the approach of soft constraints. The design for $k=1$ is in Fig.~\ref{fig:holography_penalty}C, and the corresponding $|E|^2$ obtained from FDFD is in Fig.~\ref{fig:holography_penalty}D. In terms of the computational cost, although the penalty method requires 7 rounds of training, the total number of optimization iterations is $\sim$\num{5.6e4}, which is only 70\% more than that of the soft-constraint approach with $\mu_\mathcal{F}=2$ ($\sim$\num{3.3e4}), because as shown in Algorithm \ref{alg:penalty}, for the $k$-th training, we use the $(k-1)$-th solution as the initial guess and thus the network can be trained much faster than the first round of training which is initialized randomly.

Next, we investigate the effects of $\mu_\mathcal{F}^0$ and $\beta_\mathcal{F}$. We choose $\mu_\mathcal{F}^0=1$ and compare different values of $\beta_F$. For different $\beta_F$, we increase $\mu_\mathcal{F}$ until $\mu_\mathcal{F}>100$, and the objective has a similar behavior, i.e., decreasing first and then increasing (Fig.~\ref{fig:holography_penalty}E). Among these $\beta_F$, the smallest objective is obtained from $\beta_\mathcal{F}=2$. When $\beta_\mathcal{F}$ increases, we need less rounds of training and thus the computational cost decreases, but if $\beta_\mathcal{F}$ is very large, the computational cost increases again because in each round of training we need more iterations (Fig.~\ref{fig:holography_penalty}F). We also investigate the effects of $\mu_\mathcal{F}^0$ by fixing $\beta_\mathcal{F}=2$, and the smallest objective is obtained by using $\mu_\mathcal{F}^0=2$. Therefore, the combination of $\mu_\mathcal{F}^0=2$ and $\beta_\mathcal{F}=2$ we presented is almost the best case in the problem. 

\subsubsection{Augmented Lagrangian method}

In the penalty method, the objective eventually worsens when $\mu_\mathcal{F}^k$ is large due to the ill-conditioning. To overcome this difficulty of convergence, we employ the augmented Lagrangian method in Algorithm~\ref{alg:auglag}. The Eq.~\eqref{eq:auglag} in this case becomes
\begin{equation*}
    \mathcal{L}^k =  \mathcal{J} + \mu_\mathcal{F}^k\mathcal{L}_{\mathcal{F}} + \frac{1}{2M}\sum_{j=1}^M \left( \lambda_{\Re,j}^k \Re\left[\mathcal{F}[\mathbf{x}_j]\right] + \lambda_{\Im,j}^k \Im\left[\mathcal{F}[\mathbf{x}_j]\right] \right),
\end{equation*}
and
\begin{equation} \label{eq:holography_lambda}
    \lambda_{\Re,j}^k = \lambda_{\Re,j}^{k-1} + 2\mu_\mathcal{F}^{k-1} \Re\left[\mathcal{F}[\mathbf{x}_j]\right], \quad
    \lambda_{\Im,j}^k = \lambda_{\Im,j}^{k-1} + 2\mu_\mathcal{F}^{k-1} \Im\left[\mathcal{F}[\mathbf{x}_j]\right].
\end{equation}

\begin{figure}[htbp]
    \centering
    \includegraphics[width=\textwidth]{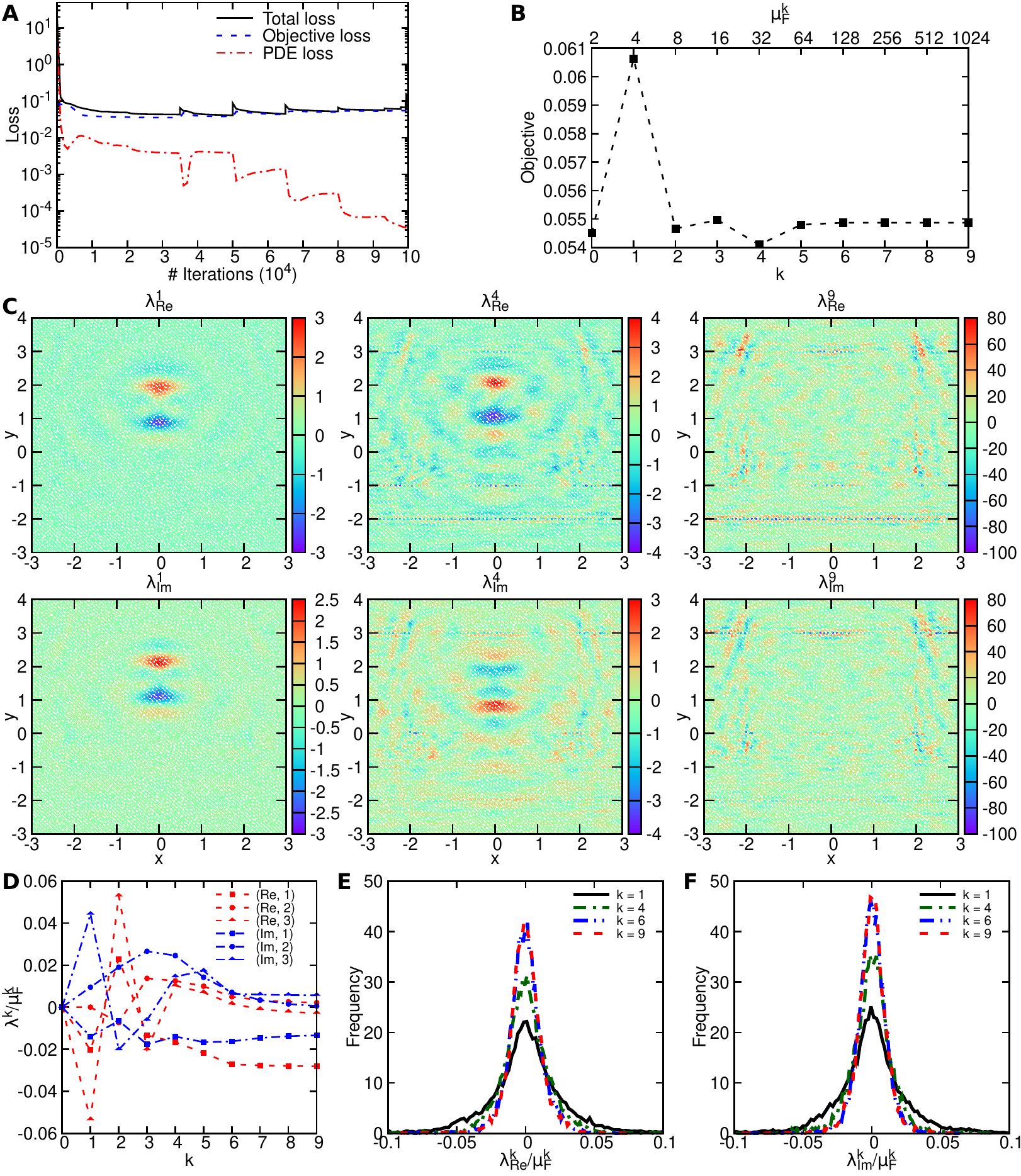}
    \caption{\textbf{hPINN for the inverse design of holography via the augmented Lagrangian method.} (\textbf{A} and \textbf{B}) The results of $\mu_\mathcal{F}^0=2$ and $\beta_\mathcal{F}=2$. (A) The losses versus the number of optimization iterations, and (B) the objective value after the network is trained with $\mu_\mathcal{F}^k$. (\textbf{C}) The values of $\lambda_{\Re,j}^k$ and $\lambda_{\Im,j}^k$ at all $\mathbf{x}_j$ locations for $k=1$, 4 and 9. (\textbf{D}) The convergence of $\lambda_{\Re,j}^k/\mu_\mathcal{F}^k$ and $\lambda_{\Im,j}^k/\mu_\mathcal{F}^k$ in three examples. (\textbf{E} and \textbf{F}) The distributions of (E) $\lambda_{\Re,j}^k/\mu_\mathcal{F}^k$ and (F) $\lambda_{\Im,j}^k/\mu_\mathcal{F}^k$ converge.}
    \label{fig:holography_auglag_ex}
\end{figure}

We use the same hyperparameters as the penalty method, i.e., $\mu_\mathcal{F}^0=2$ and $\beta_\mathcal{F}=2$. After the training, the PDE loss is below $10^{-4}$ (Fig.~\ref{fig:holography_auglag_ex}A), and the $L^2$ relative error of $|E|^2$ between hPINN and FDFD for the final $\varepsilon$ is 1.2\%, so the solution of hPINN satisfies the PDEs very well. The final value of the objective function from hPINN also . The objective does not worsen even if when $\mu_\mathcal{F}^k > 10^3$ (Fig.~\ref{fig:holography_auglag_ex}B), and thus the augmented Lagrangian method solves the convergence issue in the penalty method.

We next investigate the distribution and evolution of the  multipliers. Because we choose $\lambda_{\Re,j}^0=\lambda_{\Im,j}^0=0$ in Eq.~\eqref{eq:holography_lambda}, then $\lambda_{\Re,j}^1$ and $\lambda_{\Im,j}^1$ are proportional to the pointwise PDE loss $\mathcal{F}[\mathbf{x}_j]$. After the first round of training, the points around the region $[-0.5,0.5]\times[1,2]$ (i.e., the black square in Fig.~\ref{fig:holography}A) have the largest values of $\lambda_{\Re,j}^1$ and $\lambda_{\Im,j}^1$ (Fig.~\ref{fig:holography_auglag_ex}C left) and thus largest PDE error. Hence, the points with larger PDE errors would have larger weights for the next round of training, i.e., the multipliers automatically tune the relative weights of different training points. The distribution of $\lambda_{\Re,j}^k$ and $\lambda_{\Im,j}^k$ become more uniform for a larger $k$ (Fig.~\ref{fig:holography_auglag_ex}C). Moreover, for each single point, $\lambda_{\Re,j}^k$ and $\lambda_{\Im,j}^k$ increase with $k$, but the normalized multipliers $\lambda_{\Re,j}^k/\mu_\mathcal{F}^k$ and $\lambda_{\Im,j}^k/\mu_\mathcal{F}^k$ converges with respect to $k$, see the three examples in Fig.~\ref{fig:holography_auglag_ex}D. Also, the histogram of $\lambda_{\Re,j}^1/\mu_\mathcal{F}^1$ for all $\mathbf{x}_j$ is shown in the black curve in (Fig.~\ref{fig:holography_auglag_ex}E), which is symmetric and concentrates around zero. The distributions of $\lambda_{\Re,j}^k/\mu_\mathcal{F}^k$ converges with respect to $k$, and becomes almost converged when $k \ge 6$ (Fig.~\ref{fig:holography_auglag_ex}E). Similarly,
$\lambda_{\Im,j}^k/\mu_\mathcal{F}^k$ also converges in a similar way (Fig.~\ref{fig:holography_auglag_ex}F).

\begin{figure}[htbp]
    \centering
    \includegraphics[width=\textwidth]{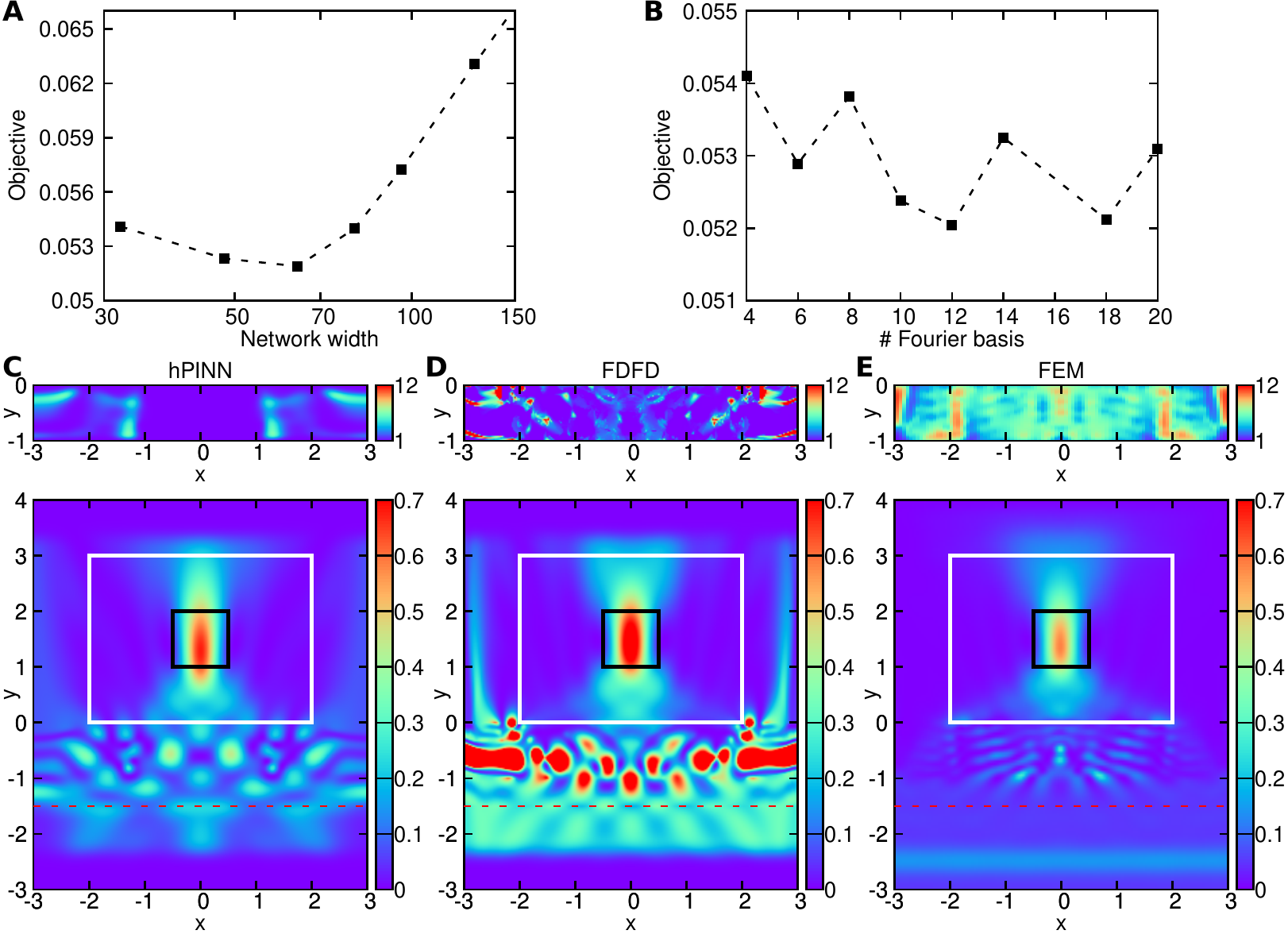}
    \caption{\textbf{Effect of hyperparameters and comparison between hPINN, FDFD, and FEM.} (\textbf{A}) Effect of the network width on the objective. (\textbf{B}) Effect of the number of Fourier basis used in the periodic BC on the objective. (\textbf{C} to \textbf{E}) The design $\varepsilon$ and the field of $|E|^2$ of (C) hPINN, (D) FDFD, and (E) FEM.}
    \label{fig:holography_hyperpar}
\end{figure}

We have demonstrated the effectiveness of the augmented Lagrangian method, and to further improve our design, we tune two hyperparameters: the network width and the number of Fourier basis used in the periodic BC. We need to choose an optimal network size to avoid the problems of underfitting and overfitting, and the optimal network width is around 60 when we choose the depth to be 5  (Fig.~\ref{fig:holography_hyperpar}A). As we discussed in Section \ref{sec:bc}, two Fourier basis is sufficient in the network to implement the periodic BC, but by using more basis terms, we can achieve smaller objective (Fig.~\ref{fig:holography_hyperpar}B). Hence, we choose the width as 48 and use 12 Fourier basis terms. This problem has many local optima with similar performance, and the exact solution found depends on the initialization of $\varepsilon$: when we trained the network from three random initializations, the objective values were 0.0517, 0.0525, and 0.0528.

As a comparison, we also solve this problem by using the method for PDE-constrained inverse design discussed in~\cite{molesky2018inverse}, and here we have considered two numerical PDE solvers: FDFD~\cite{champagne2001fdfd} and FEM~\cite{Gridap2020,bendsoe2013topology,TO2} (details in Appendices~\ref{sec:fdfd} and~\ref{sec:fem}, respectively). The objective values obtained from FDFD and FEM are 0.0523 and 0.0528, respectively, which is almost the same as that of hPINN, even though the designs are very different (because of the many local optima discussed above). In addition to the good performance, an apparent advantage of hPINN is that the design obtained from hPINN (Fig.~\ref{fig:holography_hyperpar}C top) is smoother and simpler than the designs of FDFD (Fig.~\ref{fig:holography_hyperpar}D top) and FEM (Fig.~\ref{fig:holography_hyperpar}E top). This could be explained by the analysis that neural networks trained with gradient descent have an implicit regularization and tend to converge to smooth solutions~\cite{nagarajan2019generalization,poggio2017theory,jin2020quantifying}. The electrical fields $|E|^2$ in the target domain $\Omega_3$ for these three methods are similar (Figs. \ref{fig:holography_hyperpar}C, D and E bottom).

\subsection{Fluids in Stokes flow}

Optimal design has wide and valuable applications in many fluid mechanics problems. Next, we use the proposed hPINN to solve the problem of topology optimization of fluids in Stokes flow, which was introduced by~\cite{borrvall2003topology} and has been considered as a benchmark example in many works since then~\cite{guest2006topology,duan2016topology,schulz2016computational}.

\subsubsection{Problem setup}

\begin{figure}[htbp]
    \centering
    \includegraphics[width=\textwidth]{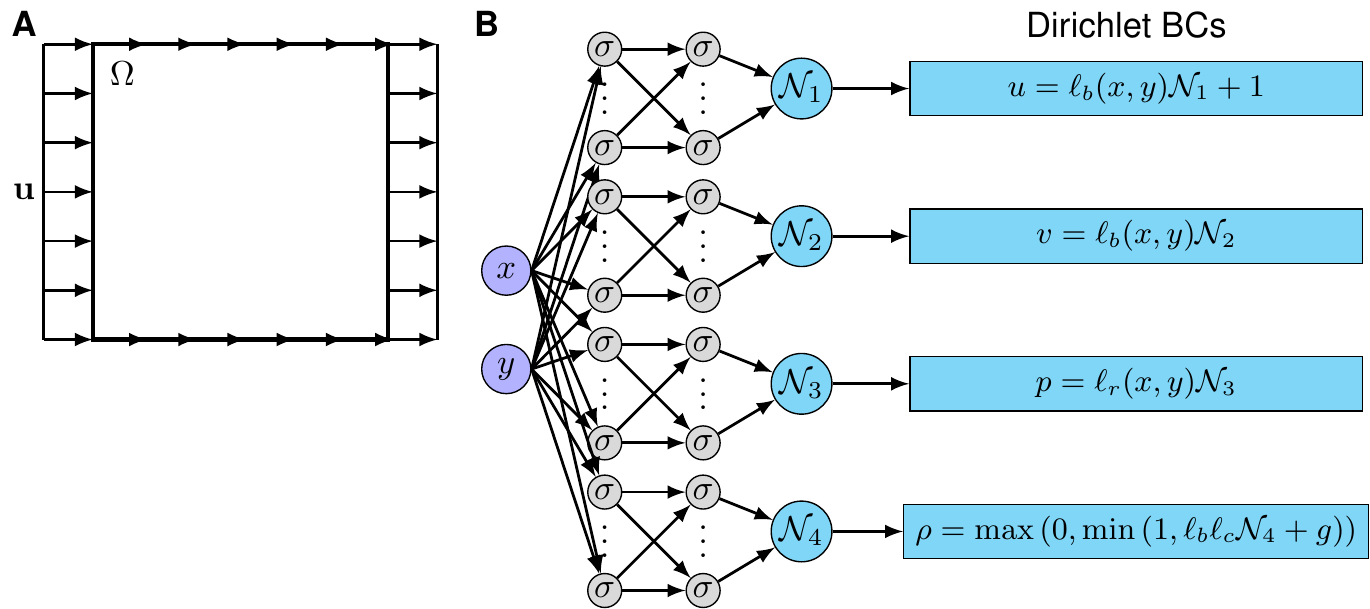}
    \caption{\textbf{Problem setup of fluids in Stokes flow and the neural network architecture.} (\textbf{A}) The design domain with the boundary condition. (\textbf{B}) The architecture of the hPINN with the Dirichlet BCs embedded directly in the network. The network inputs are $x$ and $y$, and the outputs are $u$, $v$, $p$, and $\rho$.}
    \label{fig:stokes}
\end{figure}

In this problem, we consider a design domain $\Omega = [0,1]\times[0,1]$ composed of solid material and fluids (Fig.~\ref{fig:stokes}A). The goal is to determine at what places of $\Omega$ there should be fluid and where there should be solid, in order to minimize an objective function of dissipated power. We use $\rho = 0$ to represent the solid places, and $\rho=1$ as the fluid. Then this problem is a discrete topology optimization for $\rho$, and solving a large-scale discrete topology optimization is generally computationally prohibitive~\cite{guest2006topology}. To circumvented this issue, the common procedure is to allow the intermediate media values of $\rho$ between 0 and 1.

We consider the flow is a Stokes flow, and the fluid and the solid satisfy a generalized Stokes equation by treating the solid phase as a porous medium with flow governed by the Darcy's law~\cite{borrvall2003topology,guest2006topology,duan2016topology}:
\begin{gather}
    -\nu \Delta \mathbf{u} + \nabla p = \mathbf{f}, \label{eq:stokes_f} \\
    \nabla \cdot \mathbf{u} = 0, \label{eq:stokes_u}
\end{gather}
where $\mathbf{u}=(u,v)$ is the velocity, and $p$ is the pressure. $\nu=1$ is the viscosity, and $\mathbf{f} = \alpha \mathbf{u}$ is the Brinkman term from the Darcy's law. $\alpha$ is the inverted permeability depending on $\rho$, and we use the following interpolation function proposed in~\cite{borrvall2003topology}
\begin{equation*}
    \alpha(\rho) = \overline{\alpha} + (\underline{\alpha} - \overline{\alpha}) \rho \frac{1+q}{\rho+q},
\end{equation*}
where $\overline{\alpha} = \frac{2.5\nu}{0.01^2}$ and $\underline{\alpha} = 0$ are the inverted permeability of solid and fluid phases, respectively. The parameter $q > 0$ is used to control the transition between solid and fluid phases. When $q$ is large, the interpolation has a sharper transition, and the optimization becomes more ill-conditioned. Here, we choose $q=0.1$ as suggested in~\cite{borrvall2003topology}. At the boundary, the velocity is constant $\mathbf{u} = (1, 0)$, and the pressure at the right boundary is zero (Fig.~\ref{fig:stokes}A).

The objective function of dissipated power is defined as
\begin{equation} \label{eq:stokes_obj}
    \mathcal{J} = \int_\Omega \left( \frac{1}{2}\nabla \mathbf{u}:\nabla\mathbf{u} + \frac{1}{2}\alpha\mathbf{u}^2 \right)dxdy.
\end{equation}
To make the problem meaningful, we also need to consider a fluid volume constraint:
\begin{equation*}
    \int_\Omega \rho \, dxdy \le \gamma,
\end{equation*}
and the volume fraction $\gamma$ is chosen as 0.9. Without this volume constraint, the optimal solution is $\rho=1$ everywhere.

\subsubsection{hPINN}

The PDE-informed loss function for this problem is
\begin{equation*}
    \mathcal{L}_\mathcal{F} = \frac{1}{3M}\sum_{j=1}^M (\mathcal{F}_1[\mathbf{x}_j])^2 + (\mathcal{F}_2[\mathbf{x}_j])^2 + (\mathcal{F}_3[\mathbf{x}_j])^2,
\end{equation*}
where $\mathcal{F}_1$ and $\mathcal{F}_2$ correspond to the $x$ and $y$ components of Eq.~\eqref{eq:stokes_f}, respectively, and $\mathcal{F}_3$ corresponds to Eq.~\eqref{eq:stokes_u}. Similar to the holography problem, we scale $\mathcal{F}_1$ and $\mathcal{F}_2$ by 0.01 and $\mathcal{F}_3$ by 100. In this problem, we also have an inequality constraint for the fluid volume:
\begin{equation*}
    h(\rho) = \int_\Omega \rho \, dxdy - \gamma \le 0,
\end{equation*}
which is used to compute the loss in Eqs.~\eqref{eq:loss_h} and~\eqref{eq:auglag}.

We construct four networks to approximate $u$, $v$, $p$, and $\rho$ (Fig.~\ref{fig:stokes}B), and each network has 5 layers with 64 neurons per layer. To impose the Dirichlet BCs of $u$ and $v$ into the network, we choose
\begin{equation*}
    \ell_b(x,y) = 16xy(1 - x)(1 - y),
\end{equation*}
which is equal to zero at the boundary, and the coefficient ``16'' is used to scale the function to be of order one inside $\Omega$. Similarly, for the BC of $p$ at the right boundary, we use
\begin{equation*}
    \ell_r(x,y) = 1-x.
\end{equation*}
To restrict $\rho$ between 0 and 1, we use the function $\max(0, \min(1, \cdot))$. We do not use the sigmoid function as in the holography problem, because the sigmoid function cannot be equal to 0 and 1 exactly. To prevent the optimization getting stuck at a local minimum, in~\cite{borrvall2003topology,duan2016topology}, a small portion of $\Omega$ close to the boundary is prescribed as fluid, and the initial guess of $\rho$ should be chosen properly. Here, we also add similar restrictions: the boundary is prescribed as fluid, and the center is prescribed as solid, i.e., $\rho(x,y) = 1$ when $(x,y)$ is at the boundary, and $\rho(x,y)=0$ when $x=y=0.5$. This is imposed by constructing $\rho$ as
\begin{equation} \label{eq:rho}
    \rho(x,y) = \max\left(0, \min\left(1, \ell_b(x,y)\ell_c(x,y)\mathcal{N}_4(x,y) + g(x,y)\right)\right),
\end{equation}
where
\begin{gather*}
    \ell_c(x,y) = (x-0.5)^2 + (y-0.5)^2, \\
    g(x,y) = \left(1 + \frac{\epsilon}{\min_{(x,y)} \ell_c(x,y)}\right) \left(1-\ell_b(x,y)\right) \frac{\ell_c(x,y)}{\ell_c(x,y)+\epsilon}.
\end{gather*}
By choosing a small number $\epsilon > 0$ (e.g., $\epsilon=10^{-6}$), it is easy to check that $\rho(0.5,0.5)=0$ and $\rho(x,y) = 1$ when $(x,y)$ is at the boundary.

\subsubsection{Soft constraints}

We first solve this problem using the approach of soft constraints by minimizing Eq.~\eqref{eq:soft} with $\mu_\mathcal{F} = 0.1$ and $\mu_h = 10^4$. We use 10000 points for training, i.e., $M=10000$. The same training procedure in the holography problem is used, but with a smaller learning rate \num{1e-4}.

\begin{figure}[htbp]
    \centering
    \includegraphics[width=\textwidth]{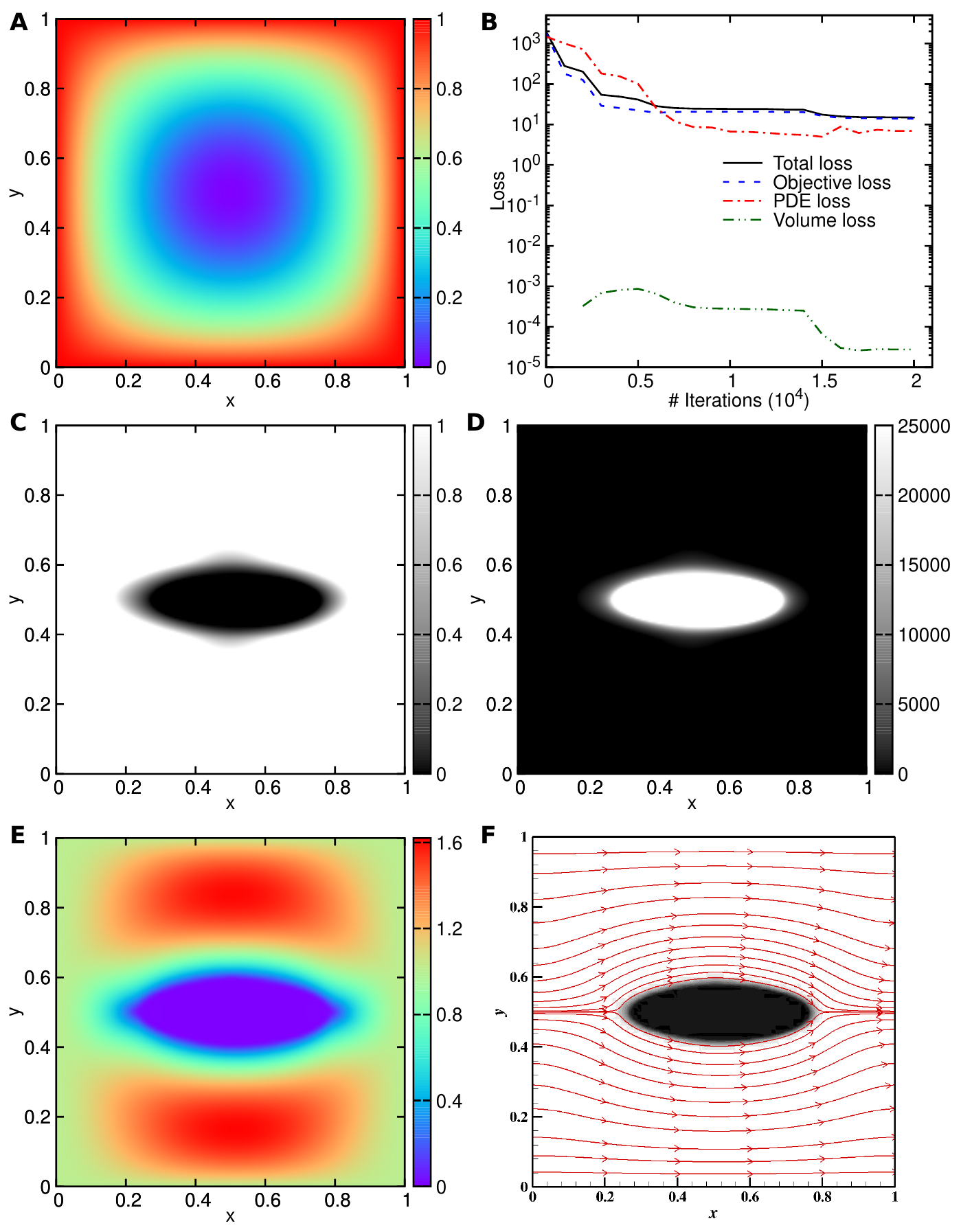}
    \caption{\textbf{hPIN for the topology optimization of fluids in Stokes flow via the soft constraint approach.} (\textbf{A}) The initialization of $\rho$. (\textbf{B}) The training losses versus the number of optimization iterations. (\textbf{C} to \textbf{F}) The optimization result: (C) $\rho$, (D) inverted permeability $\alpha$, (E) velocity magnitude $|\mathbf{u}|$, and (F) the velocity streamline.}
    \label{fig:stokes_soft}
\end{figure}

One random initialization of $\rho$ using Eq.~\eqref{eq:rho} is shown in Fig.~\ref{fig:stokes_soft}A, which satisfies our requirements for $\rho$ at the boundary and center. Using this initial guess, the loss trajectory during the training is shown in Fig.~\ref{fig:stokes_soft}B. Initially, $\rho$ satisfies the volume constraint, and after about 2000 iterations, the fluid volume $\int_\Omega \rho \, dxdy$ is always larger than $\gamma$, because more fluid makes the objective smaller. At the end of the training, the fluid volume is about 0.906, which is 0.7\% large than $\gamma$. The final fields of $\rho$ (Fig.~\ref{fig:stokes_soft}C) and $\alpha$ (Fig.~\ref{fig:stokes_soft}D) have rugby ball-like shapes, which is consistent with the designs optimized by other methods~\cite{borrvall2003topology,guest2006topology,duan2016topology,schulz2016computational}. We use the smoothed profile method (SPM)~\cite{WANG201884,LUO20091750} in the context of the spectral element method~\cite{karniadakis2005spectral} to compute the velocity from the design (see the details in Section \ref{sec:spm}), and the velocity magnitude $|\mathbf{u}|$ and the streamline are in Figs. \ref{fig:stokes_soft}E and F. We note that although the PDE loss only decreases by two orders of magnitudes from $\sim$\num{1.5e3} to $\sim$7 (Fig.~\ref{fig:stokes_soft}B), the $L^2$ relative error of $|\mathbf{u}|$ between hPINN and SPM is 4.3\%, i.e., the PDEs are satisfied well.

The objective value in Eq.~\eqref{eq:stokes_obj} of our design is 13.73. For comparison, in~\cite{borrvall2003topology}, the same setup and parameters are used as ours, and their objective value obtained from the method of moving asymptotes with FEM is 14.07, which is 2.4\% worse than ours. However, this does not mean that the soft-constraints approach is a better method: our objective is smaller only because the design slightly violates the volume constraint, permitting a reduced objective value.

\subsubsection{Augmented Lagrangian method}

\begin{figure}[htbp]
    \centering
    \includegraphics[width=\textwidth]{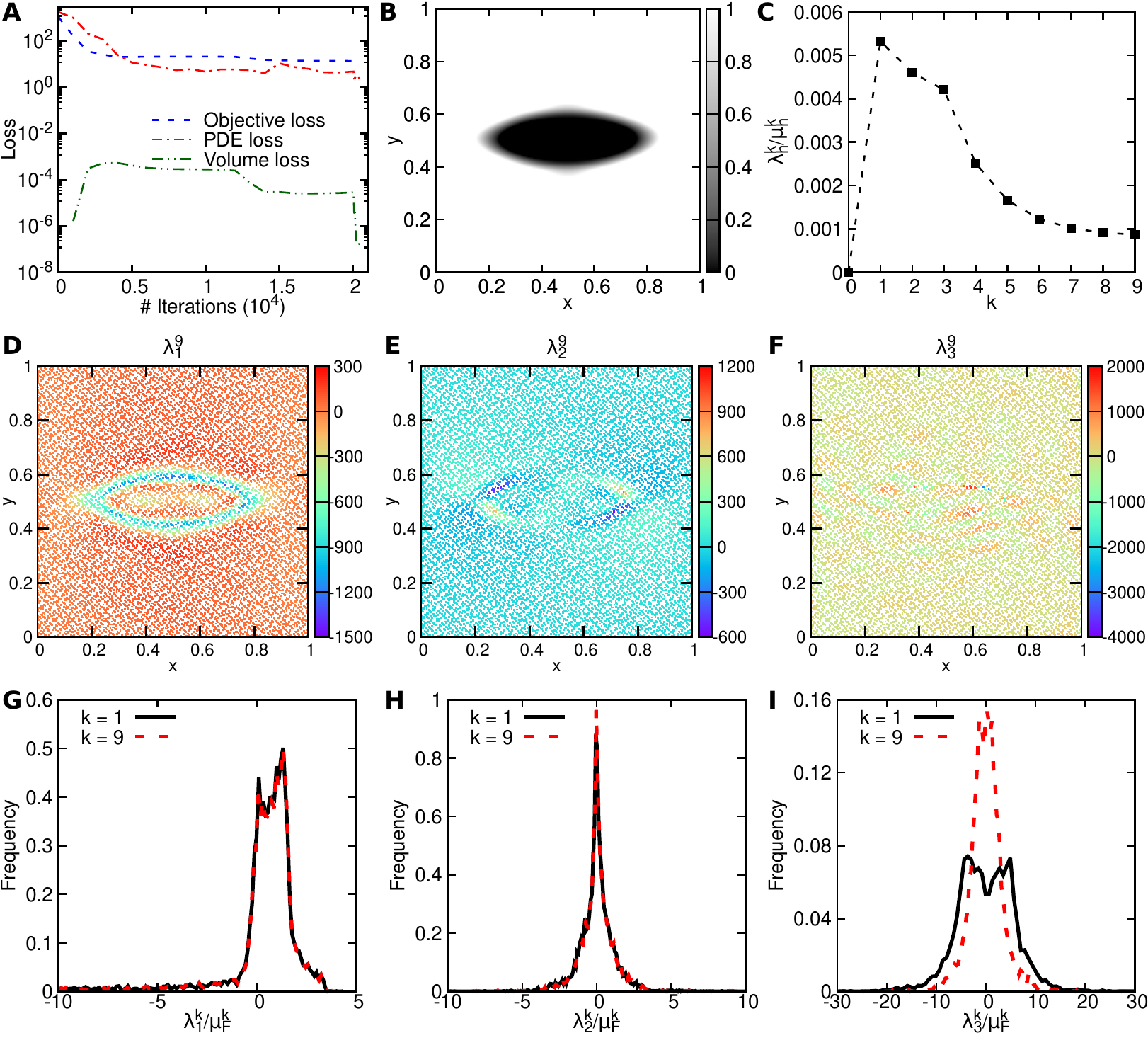}
    \caption{\textbf{hPIN for the topology optimization of fluids in Stokes flow via the augmented Lagrangian method.} (\textbf{A}) The training losses versus the number of optimization iterations. (\textbf{B}) The design of $\rho$. (\textbf{C}) $\lambda_h^k/\mu_h^k$ converges with respect to $k$. (\textbf{D} to \textbf{F}) The values of (D) $\lambda_{1,j}^9$, (E) $\lambda_{2,j}^9$, and (F) $\lambda_{3,j}^9$ at all $\mathbf{x}_j$. (\textbf{G} to \textbf{I}) The distributions of (G) $\lambda_{1,j}^k/\mu_\mathcal{F}^k$, (H) $\lambda_{2,j}^k/\mu_\mathcal{F}^k$, and (I) $\lambda_{3,j}^k/\mu_\mathcal{F}^k$ for $k=1$ and $k=9$.}
    \label{fig:stokes_auglag}
\end{figure}

To make the design satisfy the constraints better, we use the augmented Lagrangian method with $\mu_\mathcal{F}^0 = 0.1$, $\mu_h^0 = 10^4$, $\beta_\mathcal{F}=\beta_h = 2$, and $k \le 9$. As we discussed in the previous section, in the soft constraint approach, the network solution has already satisfied the PDE and volume constraints well, and thus when using the augmented Lagrangian method, we only need $\sim 400$ more iterations (Fig.~\ref{fig:stokes_soft}B and Fig.~\ref{fig:stokes_auglag}A), i.e., $\sim 2\%$ more computational cost.

After training with the augmented Lagrangian method, compared to the soft constraint approach, the loss of the fluid volume decreases from $10^{-5}$ to $10^{-8}$ (Fig.~\ref{fig:stokes_auglag}A). The final design (Fig.~\ref{fig:stokes_auglag}B) has the fluid volume of 0.901, which is almost identical to our restriction $\gamma$. On the other hand, the PDE loss only becomes a little bit smaller, and the $L^2$ relative error of $|\mathbf{u}|$ between hPINN and SPM is improved to 2.0\%. The corresponding objective value is 14.12, which is consistent with the result in~\cite{borrvall2003topology}.

Similar to the holography problem, here the normalized multipliers also converge. The normalized multiplier for the volume constraint $\lambda_h^k/\mu_h^k$ converges to 0.001 (Fig.~\ref{fig:stokes_auglag}C). The values of $\lambda_{i,j}^k$ for $i=1,2$, and 3 at $k=9$ are shown in Figs. \ref{fig:stokes_auglag}D, E, and F, respectively. $\lambda_{3,j}^9$ is almost uniform, but $\lambda_{1,j}^9$ and $\lambda_{2,j}^9$ have the largest values near the interface between the solid and fluid phases. The distributions of normalized $\lambda_{1,j}^k$ and $\lambda_{2,j}^k$ do not change too much from $k=1$ to $k=9$ (Figs. \ref{fig:stokes_auglag}G and H). However, the distribution of normalized $\lambda_{3,j}^k$ changes a lot and converges to a Gaussian distribution (Fig.~\ref{fig:stokes_auglag}I), because $\lambda_{3,j}^k$ corresponds to Eq.~\eqref{eq:stokes_u}, i.e., the mass continuity equation for incompressible fluid. This is consistent with our analysis that during the training, the fluid volume is the main variable to be improved.

\section{Conclusion} \label{sec:conc}

In an inverse design problem, we aim to find the best design by minimizing an objective function, which is subject to various constraints, including partial differential equations (PDEs), boundary conditions (BCs), and inequalities. We have developed in this study a new method of physics-informed neural networks (PINNs) with hard constraints (hPINNs) for solving inverse design. In hPINN, we enforce the PDE and inequality constraints by using loss functions, while we impose exactly Dirichlet and periodic BCs into the neural network architecture. We proposed the two approaches of the penalty method and the augmented Lagrangian method to enforce the loss function as hard constraints.

We demonstrated the effectiveness of hPINN for two examples: the holography problem in optics and fluids of Stokes flow. Our numerical results of both examples show that although the approach of soft constraints cannot satisfy the PDE or inequality constraints to a good accuracy during the network training, we may still obtain a relatively good design, if the penalty coefficients are chosen properly. The penalty method are able to impose hard constraints, but it has the issue of convergence when the penalty coefficients are too large. By using the augmented Lagrangian method, we can impose hard constraints and also achieve a better design. Compared to traditional PDE-constrained optimization methods based on adjoint methods and numerical PDE solvers, the results obtained from hPINN have the same objective value, but the design is simpler and smoother for the problem of non-unique solutions. In addition, the objective function and the multipliers converge quickly after only several rounds of training. We also show that the computational cost of augmented Lagrangian method is comparable to the soft constraint approach, because we use the network solution of the previous round as the network initialization for the next round of training. Moreover, we note that hPINN is a purely deep learning approach, and does not rely on any numerical solver. Hence, the implementation of hPINN is ``one-size-fit-all'' for inverse design.

In our numerical results, we show that hPINN with the augmented Lagrangian converges to a good solution, but we do not have a theoretical understanding yet. We note that the convergence of stochastic optimization with the augmented Lagrangian method has been analyzed in \cite{wang2008stochastic}, and the convergence of PINN for certain linear PDEs has been proved in \cite{shin2020convergence}. In future work, we will analyze theoretically the convergence of hPINN based on these results. In addition, in the augmented Lagrangian method, we used one multiplier for each residual location $\mathbf{x}$, and thus many multipliers are required, which could be expensive and inefficient for large-scale problems.
% This also prevents us to change training points during the optimization.
In fact, the profile of multipliers is smooth with respect to $\mathbf{x}$, and thus it may be possible to use one single multiplier function $\lambda(\mathbf{x})$ (e.g., represented by a neural network) for the entire computational domain. This approach also leverages the continuity of the multiplier, and may induce a faster convergence.

% \section*{Acknowledgments}

% This work was supported by the MIT-IBM Watson AI Laboratory (challenge 2415). F.V. acknowledges support from the Spanish Ministry of Economy and Competitiveness through the ``Severo Ochoa Programme for Centers of Excellence in R\&D (CEX2018-000797-S)''.

% \bibliographystyle{unsrt}
\bibliographystyle{siamplain}
\bibliography{main}

\clearpage
\appendix

\input{SI}

\end{document}

%% file: shared.tex
\usepackage{lipsum}
\usepackage{amsfonts}
\usepackage{graphicx}
\usepackage{epstopdf}
\usepackage{algorithmic}
\ifpdf
  \DeclareGraphicsExtensions{.eps,.pdf,.png,.jpg}
\else
  \DeclareGraphicsExtensions{.eps}
\fi

\newsiamremark{remark}{Remark}
\newsiamremark{hypothesis}{Hypothesis}
\crefname{hypothesis}{Hypothesis}{Hypotheses}
\newsiamthm{claim}{Claim}

\headers{Hard-constrained Physics-Informed Neural Network}{L. Lu, R. Pestourie, W. Yao, Z. Wang, F. Verdugo, and S. G. Johnson}

\title{Physics-informed neural networks with hard constraints for inverse design\thanks{Submitted to the editors DATE.
\funding{This work was supported by the MIT-IBM Watson AI Laboratory (challenge 2415). F.V. acknowledges support from the Spanish Ministry of Economy and Competitiveness through the ``Severo Ochoa Programme for Centers of Excellence in R\&D (CEX2018-000797-S)''.}}}

\author{Lu Lu\thanks{Department of Mathematics, Massachusetts Institute of Technology, Cambridge, MA 02139, USA (\email{lu\_lu@mit.edu}, \email{stevenj@math.mit.edu}).}
\and Rapha\"el Pestourie\footnotemark[2]
\and Wenjie Yao\footnotemark[2]
\and Zhicheng Wang\thanks{Laboratory of Ocean Energy Utilization of Ministry of Education, Dalian University of Technology, Dalian, 116024, China}
\and Francesc Verdugo\thanks{Centre Internacional de M\`etodes Num\`erics en Enginyeria, Esteve Terradas 5, 08860 Castelldefels, Spain}
\and Steven G. Johnson\footnotemark[2]
}

\usepackage{amsopn}

%% file: SI.tex
\section{PDEs of the holography problem} \label{sec:holography_pde}

The real and imaginary parts of Eq. (3.3)
% \eqref{eq:pml}
are:
\[\left(A_1\frac{\partial^2}{\partial x^2} + A_2\frac{\partial}{\partial x} + A_3\frac{\partial^2}{\partial y^2} + A_4\frac{\partial}{\partial y}\right)\Re[E] - \left(B_1\frac{\partial^2}{\partial x^2} + B_2\frac{\partial}{\partial x} + B_3\frac{\partial^2}{\partial y^2} + B_4\frac{\partial}{\partial y}\right)\Im[E] + \varepsilon\omega^2\Re[E] = 0,\]
\[\left(A_1\frac{\partial^2}{\partial x^2} + A_2\frac{\partial}{\partial x} + A_3\frac{\partial^2}{\partial y^2} + A_4\frac{\partial}{\partial y}\right)\Im[E] + \left(B_1\frac{\partial^2}{\partial x^2} + B_2\frac{\partial}{\partial x} + B_3\frac{\partial^2}{\partial y^2} + B_4\frac{\partial}{\partial y}\right)\Re[E] + \varepsilon\omega^2\Im[E] = -\omega J,\]
where
\[\frac{1}{(1+i\frac{\sigma_x}{\omega})^2} = A_1(x) + iB_1(x),\]
\[-\frac{i\frac{\sigma_x'}{\omega}}{(1+i\frac{\sigma_x}{\omega})^3} = A_2(x) + iB_2(x),\]
\[\frac{1}{(1+i\frac{\sigma_y}{\omega})^2} = A_3(y) + iB_3(y),\]
\[-\frac{i\frac{\sigma_y'}{\omega}}{(1+i\frac{\sigma_y}{\omega})^3} = A_4(y) + iB_4(y),\]
and
$$\sigma'_x(x)=-2\sigma_0(-2-x) \mathbbm{1}_{(-\infty, -2)}(x) + 2\sigma_0(x-2) \mathbbm{1}_{(2,\infty)}(x),$$
$$\sigma'_y(y)=-2\sigma_0(-2-y) \mathbbm{1}_{(-\infty, -2)}(y) + 2\sigma_0(y-3) \mathbbm{1}_{(3,\infty)}(y).$$

\section{Solving holography problem via the finite-difference frequency-domain method} \label{sec:fdfd}

We optimized the permittivity in the design volume to best match the targeted pattern. The PDE constraints and the adjoint method were computed via a finite-difference frequency solver as in~\cite{molesky2018inverse}. We used a resolution of 40 pixel per wavelength in vacuum for the computational domain, which corresponds to about 10 pixels per wavelength in the worst case scenario (when the permittivity is 12). We optimized the permittivity region by minimizing the objective function using the method of moving asymptotes (MMA) based on conservative convex separable approximations (CCSA)~\cite{svanberg2002class}. 

For each pixel in the design region, the minimum feasible permittivity is 1 (permittivity of the vacuum) and the maximum feasible permittivity is 12. The initial guess for the permittivity region was uniform of unit permittivity. We noticed the lower the starting permittivity was, the smoother the end design. Setting the relative tolerance in the input of the optimization algorithm to $10^{-8}$, the opitmization converged in about 5000 steps.
% for about 2h on an Intel Core i7-8559U processor.
Notice that a more judicious choice of optimization algorithm parameters might has led to less optimization steps. However, since the optimization converged, the optimum is representative of the method performance.

\section{Solving holography problem via the finite element method} \label{sec:fem}

In the finite element method, we use different mesh resolutions in different regions. The resolution is 20 pixel per wavelength in the $\epsilon=1$ domain, while the resolution doubles in the design domain and half in the PML. The computation domain is meshed by GMSH~\cite{gmsh} and then the PDEs are solved via Gridap.jl~\cite{Gridap2020}.  

The optimization is done via the traditional topology optimization techniques~\cite{bendsoe2013topology,TO2}. For each pixel in the design region, the permittivity is characterized by $\epsilon=1+11p$, where $p\in[0,1]$ is the design parameter. These design parameters are further smoothed by a Helmholtz-type filter~\cite{TOfilter} with a filter radius about double the minimum mesh length scale. The derivative to these design parameters are obtained by the adjoint method, and then we use the method of moving asymptotes (MMA) based on conservative convex separable approximations (CCSA)~\cite{svanberg2002class} to optimize these parameters.  For initial guess, $p$ is chosen to be $0.5$ for all pixels. The stopping criteria is set to be a $10^{-6}$ relative difference in the loss function, it took about 500 iterations
% (about 15 minutes on an Intel Core i7-8700K processor)
for the optimization to converge. We note that with a different initial guess, the optimization might converge to different shapes, indicating that there are multiple local optima in this optimization problem.

\section{Smoothed profile method for Stokes flow} \label{sec:spm}

In general, in the smoothed profile method \cite{WANG201884}, the immersed solid body is represented by the following hyperbolic tangent function,
%%%%%%%%%%%%%%%%%%%%%%%%%%%%%%%%%%%%%%%%%%%%%%%%%%%%%%%%%%%%%%%%%%%%%%%%%%%%%%%%%%%%%%%%%%%%%%%%%%%%
\begin{equation*}
 \phi (\mathbf{x})=\frac{1}{2}\left[\mathrm{tanh}\left(\frac{-d(\mathbf{x})}{\xi}\right)+1\right] ,
\end{equation*}
%%%%%%%%%%%%%%%%%%%%%%%%%%%%%%%%%%%%%%%%%%%%%%%%%%%%%%%%%%%%%%%%%%%%%%%%%%%%%%%%%%%%%%%%%%%%%%%%%%%%%%
where $d(\mathbf{x}) $ is the signed distance to surface of the immersed body, $\xi$ is the interface thickness parameter, and  $\phi (\mathbf{x})$ is a function of spatial coordinates $\mathbf{x}$; it is equal to $1$ inside the riser, $0$ in the fluid domain, and varies smoothly between $1$ and $0$ in the solid-fluid interfacial layer.  However, in current study, $\phi(\mathbf{x})$ is obtained from the inverse permeability parameter $\alpha$, which is re-scaled to the range $[0,1]$.

The fluid flow is governed by the Stokes equations:
%%%%%%%%%%%%%%%%%%%%%%%%%%%%%%%%%%%%%%%%%%%%%%%%%%%%%%%%%%%%%%%%%%%%%%%%%%%%%%%%%%%%%%%%%%%%%%%%%%%%%
\begin{equation}\label{eq:SPM_NSE1}
   \nabla \cdot \mathbf{u} = 0,
\end{equation}
\begin{equation}\label{eq:SPM_NSE2}
   \frac{\partial \mathbf{u}}{\partial t} =-\nabla p+\nu \nabla^{2} \mathbf{u} .
\end{equation}
%%%%%%%%%%%%%%%%%%%%%%%%%%%%%%%%%%%%%%%%%%%%%%%%%%%%%%%%%%%%%%%%%%%%%%%%%%%%%%%%%%%%%%%%%%%%%%%%%%%%%%%
Given $(\mathbf{u}^{n}, p^{n}, \phi)$, we first obtain $\hat{\mathbf{u}}$ as follows,
%%%%%%%%%%%%%%%%%%%%%%%%%%%%%%%%%%%%%%%%%%%%%%%%%%%%%%%%%%%%%%%%%%%%%%%%%%%%%%%%%%%%%%%%%%%%%%%%%
\begin{equation}\label{eq:nonlinear_stage}
\hat{\mathbf{u}} = \Delta t \sum_{q=0}^{J-1} \alpha_q \mathbf{u} ^{n-q},
\end{equation}
%%%%%%%%%%%%%%%%%%%%%%%%%%%%%%%%%%%%%%%%%%%%%%%%%%%%%%%%%%%%%%%%%%%%%%%%%%%%%%%%%%%%%%%%%%%%%%%%%%%%%
where $\alpha_q$ is the coefficient of the stiffly-stable integration scheme we employ with $J=2$ the integration order. 
In the next stage we solve the intermediate pressure field,
\begin{equation}
\nabla^2 p^{*}=\nabla \cdot \left(\frac{\hat{\mathbf{u}}}{\Delta t}\right),
\end{equation} 
with the following pressure boundary condition at all the velocity Dirichlet boundaries,
\begin{equation}
\frac{\partial p^{*}}{\partial \mathbf{n}}=\sum_{q=0}^{J-1}[-\nu \nabla \times (\nabla \times \mathbf{u})]^{n-q} \cdot  \mathbf{n},
\end{equation}
where $ \mathbf{n}$ is the unit outward normal vector at the boundaries.

In the third stage of the method we compute the intermediate velocity $\mathbf{u}^{*}$,
\begin{equation}
\left(\nabla^2-\frac{\gamma_0}{\nu \Delta t}\right) \mathbf{u}^{*}=-\frac{\hat{\mathbf{u}}}{\nu \Delta t},
\end{equation}
where $\gamma_0$ is the scaled coefficient of the stiffly-stabled scheme \cite{karniadakis2005spectral}. 
\par
If this is the first iteration, then the fourth stage to obtain the immersed body velocity is as follows,
\begin{equation}
 \mathbf{u}_{p}=\phi \mathbf{V}_{s},
\end{equation}
where $\mathbf{V}_{s}$ is the velocity of the immersed bluff-body, here we have used $\mathbf{V}_{s}=0$.  

Next, we solve the extra pressure field $p_p$ due to the immersed solid body,
\begin{equation}
\nabla^2 p_{p}=\nabla \cdot \left(\frac{\gamma_0 \phi (\mathbf{u}_{p}-\mathbf{u}^{*})}{\Delta t}\right).
\end{equation}
Here the following is used as the boundary conditions for $p_p$ at any velocity Dirichlet boundary,
\begin{equation}
\frac{\partial p_{p}}{\partial \mathbf{n}}= \frac{\gamma_0 \phi (\mathbf{u}_{p}-\mathbf{u}^{*})}{\Delta t} \cdot \mathbf{n}.
\end{equation}
Finally, the total velocity field and pressure are updated as follows,
\begin{equation}
\frac{\gamma_0 \mathbf{u}^{n+1}-\gamma_0 \mathbf{u}^{*}}{\Delta t}=\frac{\gamma_0 \phi (\mathbf{u}_{p}-\mathbf{u}^{*})}{\Delta t}-\nabla p_{p},
\end{equation}
\begin{equation} \label{eq:final_stage}
p^{n+1}=p^{*}+p_p.
\end{equation}
Note that through Eqs. \eqref{eq:nonlinear_stage}--\eqref{eq:final_stage}, the no-slip and no-penetration boundary conditions are fulfilled automatically \cite{LUO20091750}.  

In all the SPM simulations of this paper, a computational domain of size $[0,1] \times [0,1]$ that consists of $50 \times 50$ uniform quadrilateral elements is used. The spectral element method \cite{karniadakis2005spectral} with 2nd order polynomials and a time step of $\Delta t=10^{-4}$ are employed to integrate the governing equations \eqref{eq:SPM_NSE1} and \eqref{eq:SPM_NSE2} until steady state.